\RequirePackage{lineno} 
\documentclass[longbibliography,aps,prc,superscriptaddress,twocolumn,nofootinbib,amsmath,amssymb,showpacs,showkeys,floatfix,preprintnumbers]{revtex4-1}

\usepackage{pifont}
\usepackage{graphicx}
\usepackage{color}
\usepackage{amsmath}
\usepackage{amssymb}

\usepackage[caption=false]{subfig}
\usepackage{multirow}

\usepackage{longtable}
\usepackage[percent]{overpic}

\graphicspath{ {figures/} }

\begin{document} 



\title{Beam-Target Helicity Asymmetry $E$ in $K^{0}\Lambda$ and $K^{0}\Sigma^0$ \\ Photoproduction on the Neutron}

\newcommand*{\CMU}{Carnegie Mellon University, Pittsburgh, Pennsylvania 15213}
\newcommand*{\CMUindex}{1}
\affiliation{\CMU}
\newcommand*{\ANL}{Argonne National Laboratory, Argonne, Illinois 60439}
\newcommand*{\ANLindex}{2}
\affiliation{\ANL}
\newcommand*{\ASU}{Arizona State University, Tempe, Arizona 85287-1504}
\newcommand*{\ASUindex}{3}
\affiliation{\ASU}
\newcommand*{\CSUDH}{California State University, Dominguez Hills, Carson, CA 90747}
\newcommand*{\CSUDHindex}{4}
\affiliation{\CSUDH}
\newcommand*{\CANISIUS}{Canisius College, Buffalo, NY, 14208}
\newcommand*{\CANISIUSindex}{5}
\affiliation{\CANISIUS}
\newcommand*{\CUA}{Catholic University of America, Washington, D.C. 20064}
\newcommand*{\CUAindex}{6}
\affiliation{\CUA}
\newcommand*{\SACLAY}{IRFU, CEA, Universit\'e Paris-Saclay, F-91191 Gif-sur-Yvette, France}
\newcommand*{\SACLAYindex}{7}
\affiliation{\SACLAY}
\newcommand*{\CF}{Univesit\'e Blaise Pascal, Clermont-Ferrand, Aubi\`ere Cedex 63178, France}
\newcommand*{\CFindex}{7}
\affiliation{\CF}
\newcommand*{\CNU}{Christopher Newport University, Newport News, Virginia 23606}
\newcommand*{\CNUindex}{8}
\affiliation{\CNU}
\newcommand*{\UCONN}{University of Connecticut, Storrs, Connecticut 06269}
\newcommand*{\UCONNindex}{9}
\affiliation{\UCONN}
\newcommand*{\DUKE}{Duke University, Durham, North Carolina 27708-0305}
\newcommand*{\DUKEindex}{10}
\affiliation{\DUKE}
\newcommand*{\FU}{Fairfield University, Fairfield CT 06824}
\newcommand*{\FUindex}{11}
\affiliation{\FU}
\newcommand*{\FERRARAU}{Universita' di Ferrara , 44121 Ferrara, Italy}
\newcommand*{\FERRARAUindex}{12}
\affiliation{\FERRARAU}
\newcommand*{\FIU}{Florida International University, Miami, Florida 33199}
\newcommand*{\FIUindex}{13}
\affiliation{\FIU}
\newcommand*{\FSU}{Florida State University, Tallahassee, Florida 32306}
\newcommand*{\FSUindex}{14}
\affiliation{\FSU}
\newcommand*{\GACHINA}{Particle and Nuclear Physics Institute, Orlova Rosha 1, 188300 Gatchina, Russia}
\newcommand*{\GACHINAindex}{45}
\affiliation{\GACHINA}
\newcommand*{\GENOVA}{Universit\`a di Genova, Dipartimento di Fisica, 16146 Genova, Italy}
\newcommand*{\GENOVAindex}{14}
\affiliation{\GENOVA}
\newcommand*{\GWUI}{The George Washington University, Washington, DC 20052}
\newcommand*{\GWUIindex}{15}
\affiliation{\GWUI}
\newcommand*{\BONN}{Helmholtz-Institute f\"ur Strahlen- und Kernphysik der Rheinischen Friedrich-Wilhelms Universit\"at, Nussallee 14-16, 53115 Bonn, Germany}
\newcommand*{\BONNindex}{43}
\affiliation{\BONN}
\newcommand*{\ISU}{Idaho State University, Pocatello, Idaho 83209}
\newcommand*{\ISUIindex}{15}
\affiliation{\ISU}
\newcommand*{\IOWA}{University of Iowa, Iowa City, IA 52242}
\newcommand*{\IOWAindex}{43}
\affiliation{\IOWA}
\newcommand*{\INFNFE}{INFN, Sezione di Ferrara, 44100 Ferrara, Italy}
\newcommand*{\INFNFEindex}{16}
\affiliation{\INFNFE}
\newcommand*{\INFNFR}{INFN, Laboratori Nazionali di Frascati, 00044 Frascati, Italy}
\newcommand*{\INFNFRindex}{17}
\affiliation{\INFNFR}
\newcommand*{\INFNGE}{INFN, Sezione di Genova, 16146 Genova, Italy}
\newcommand*{\INFNGEindex}{18}
\affiliation{\INFNGE}
\newcommand*{\INFNRO}{INFN, Sezione di Roma Tor Vergata, 00133 Rome, Italy}
\newcommand*{\INFNROindex}{19}
\affiliation{\INFNRO}
\newcommand*{\INFNTUR}{INFN, Sezione di Torino, 10125 Torino, Italy}
\newcommand*{\INFNTURindex}{20}
\affiliation{\INFNTUR}
\newcommand*{\ORSAY}{Institut de Physique Nucl\'eaire, CNRS/IN2P3 and Universit\'e Paris Sud, Orsay, France}
\newcommand*{\ORSAYindex}{21}
\affiliation{\ORSAY}
\newcommand*{\ITEP}{Institute of Theoretical and Experimental Physics, Moscow, 117259, Russia}
\newcommand*{\ITEPindex}{22}
\affiliation{\ITEP}
\newcommand*{\JMU}{James Madison University, Harrisonburg, Virginia 22807}
\newcommand*{\JMUindex}{23}
\affiliation{\JMU}
\newcommand*{\KNU}{Kyungpook National University, Daegu 41566, Republic of Korea}
\newcommand*{\KNUindex}{24}
\affiliation{\KNU}
\newcommand*{\MISS}{Mississippi State University, Mississippi State, MS 39762-5167}
\newcommand*{\MISSindex}{25}
\affiliation{\MISS}
\newcommand*{\UNH}{University of New Hampshire, Durham, New Hampshire 03824-3568}
\newcommand*{\UNHindex}{26}
\affiliation{\UNH}
\newcommand*{\NSU}{Norfolk State University, Norfolk, Virginia 23504}
\newcommand*{\NSUindex}{27}
\affiliation{\NSU}
\newcommand*{\OHIOU}{Ohio University, Athens, Ohio  45701}
\newcommand*{\OHIOUindex}{28}
\affiliation{\OHIOU}
\newcommand*{\ODU}{Old Dominion University, Norfolk, Virginia 23529}
\newcommand*{\ODUindex}{29}
\affiliation{\ODU}
\newcommand*{\RPI}{Rensselaer Polytechnic Institute, Troy, New York 12180-3590}
\newcommand*{\RPIindex}{30}
\affiliation{\RPI}
\newcommand*{\ROMAII}{Universita' di Roma Tor Vergata, 00133 Rome Italy}
\newcommand*{\ROMAIIindex}{31}
\affiliation{\ROMAII}
\newcommand*{\MSU}{Skobeltsyn Institute of Nuclear Physics, Lomonosov Moscow State University, 119234 Moscow, Russia}
\newcommand*{\MSUindex}{32}
\affiliation{\MSU}
\newcommand*{\SCAROLINA}{University of South Carolina, Columbia, South Carolina 29208}
\newcommand*{\SCAROLINAindex}{33}
\affiliation{\SCAROLINA}
\newcommand*{\TEMPLE}{Temple University,  Philadelphia, PA 19122 }
\newcommand*{\TEMPLEindex}{34}
\affiliation{\TEMPLE}
\newcommand*{\JLAB}{Thomas Jefferson National Accelerator Facility, Newport News, Virginia 23606}
\newcommand*{\JLABindex}{35}
\affiliation{\JLAB}
\newcommand*{\UTFSM}{Universidad T\'{e}cnica Federico Santa Mar\'{i}a, Casilla 110-V Valpara\'{i}so, Chile}
\newcommand*{\UTFSMindex}{36}
\affiliation{\UTFSM}
\newcommand*{\EDINBURGH}{Edinburgh University, Edinburgh EH9 3JZ, United Kingdom}
\newcommand*{\EDINBURGHindex}{37}
\affiliation{\EDINBURGH}
\newcommand*{\GLASGOW}{University of Glasgow, Glasgow G12 8QQ, United Kingdom}
\newcommand*{\GLASGOWindex}{38}
\affiliation{\GLASGOW}
\newcommand*{\VCU}{Virginia Commonwealth University, Richmond, VA 23220}
\newcommand*{\VCUindex}{39}
\affiliation{\VCU}
\newcommand*{\VT}{Virginia Tech, Blacksburg, Virginia   24061-0435}
\newcommand*{\VTindex}{39}
\affiliation{\VT}
\newcommand*{\VIRGINIA}{University of Virginia, Charlottesville, Virginia 22901}
\newcommand*{\VIRGINIAindex}{40}
\affiliation{\VIRGINIA}
\newcommand*{\WM}{College of William and Mary, Williamsburg, Virginia 23187-8795}
\newcommand*{\WMindex}{41}
\affiliation{\WM}
\newcommand*{\YEREVAN}{Yerevan Physics Institute, 375036 Yerevan, Armenia}
\newcommand*{\YEREVANindex}{42}
\affiliation{\YEREVAN}

\newcommand*{\RASEMAIL}{ schumacher@cmu.edu}
\newcommand*{\NOWITG}{ Investment Technology Group, Boston, MA}
\newcommand*{\NOWMISS}{ Mississippi State University, Mississippi State, MS 39762-5167}
\newcommand*{\NOWGWUI}{ The George Washington University, Washington, DC 20052}
\newcommand*{\NOWUK}{ University of Kentucky, Lexington, KY 40506}
\newcommand*{\NOWISU}{ Idaho State University, Pocatello, Idaho 83209}

\author {D.H.~Ho} 
\altaffiliation[Current address:]{\NOWITG}
\affiliation{\CMU}
\author {R.A.~Schumacher} 
\altaffiliation[Contact:]{\RASEMAIL}
\affiliation{\CMU}
\author {A.~D'Angelo} 
\affiliation{\INFNRO}
\affiliation{\ROMAII}
\author {A.~Deur} 
\affiliation{\JLAB}
\author {J.~Fleming} 
\affiliation{\EDINBURGH}
\author {C.~Hanretty} 
\affiliation{\JLAB}
\affiliation{\VIRGINIA}
\author {T.~Kageya} 
\affiliation{\JLAB}
\author {F.J.~Klein} 
\affiliation{\CUA}
\author {E.~Klempt} 
\affiliation{\BONN}
\author {M.M.~Lowry} 
\affiliation{\JLAB}
% ...

\collaboration{The CLAS Collaboration}
\noaffiliation

\date{\today}

\begin{abstract}

We report the first measurements of the $E$ beam-target helicity asymmetry  for the $\vec{\gamma} \vec{n} \to K^{0}\Lambda$, and $K^{0}\Sigma^{0}$ channels in the energy range 1.70$\leq W\leq$2.34 GeV.  The CLAS system at Jefferson Lab uses  a  circularly polarized photon beam and a target consisting of  longitudinally polarized solid molecular hydrogen deuteride  with low background contamination for the measurements.
The multivariate analysis method  boosted decision trees was used to isolate the reactions of interest.
Comparisons with  predictions from the KaonMAID, SAID, and Bonn-Gatchina models are presented.  
These results will help separate the isospin $I=0$ and $I=1$ photo-coupling transition amplitudes in pseudoscalar meson photoproduction.

\end{abstract}

\maketitle


\section{Introduction}
\label{sec:intro}

An accurate description of excited nucleons and their interaction with probes such as photons at GeV energies has remained elusive for decades.   
The standard model~\citep{Gross:1973id,Politzer:1973fx} underpins the structure of the 
nucleons and their excitations, but in the low-energy non-perturbative regime, competing semi-phenomenological models of specific
reaction dynamics are all that are available.  Present-day  
lattice QCD calculations~\citep{Edwards:2011jj, Edwards:2012fx} and quark models~\citep{Capstick:2000qj, Capstick:1998uh, Capstick:1986bm,Loring:2001kx,Glozman:1997ag, Giannini:2001kb} predict a richer baryon spectrum than experimentally observed~\citep{Patrignani:2016xqp, Klempt:2009pi,Koniuk:1979vw}: the so-called {\it missing resonance problem}.  There are theoretical approaches to the nucleon resonance spectrum that predict that some quark-model states do not exist, including models with quasi-stable diquarks~\citep{Anselmino:1992vg},
AdS/QCD string-based models~\citep{Brodsky:2006uq}, and ``molecular'' models in which some baryon resonances are dynamically generated from
the unitarized interaction among ground-state baryons and mesons~\citep{Kolomeitsev:2003kt}.
But finding such missing states may in part be an experimental problem:  high-mass nucleon resonances may couple weakly to $\pi N$ and may thus have escaped detection in the analysis of $\pi N$ elastic scattering experiments.
Further, they are wide and overlapping,  and partial-wave analysis (PWA) of reaction data for specific final states remains difficult due to channel-coupling effects and insufficient experimental constraints.
The experimental results discussed here represent one step in the direction of adding constraints to the hyperon photoproduction database, which ultimately impacts models for nucleon excitations.  

Cross-section measurements alone are not enough to constrain PWA models of meson production amplitudes.  Polarization observables related to the spins of the beam photons, target, and recoiling baryons are also needed.
Photoproduction of pseudoscalar mesons is governed by four complex amplitudes that lead to an interaction cross sections and 15 spin observables~\citep{Chew:1957tf, Barker:1975bp,Fasano:1992es, Chiang:1996em, Keaton:1996pe, Sandorfi:2010uv, Nys:2016uel}.  
A \textit{mathematically complete} experiment would require data, with negligible uncertainties, on a minimum of eight well-chosen observables at each center-of mass (c.m.) energy, $W$,  and meson polar angle, $\cos \theta_{c.m.}$. In practice, with realistically achievable uncertainties, measurements of many more are needed to select between competing partial wave solutions, and even knowledge of the sign of an asymmetry can provide valuable discrimination~\cite{Sandorfi:2010uv}. Furthermore, avoiding ambiguities in PWA solutions requires measurements of observables from each spin configuration of the three combinations of beam-target, target-recoil, and beam-recoil polarization~\cite{Sandorfi:2010uv, Nys:2016uel}.

Furthermore, while isospin $I=3/2$ transitions ($\Delta^*$ excitations) can be studied with proton target data alone, both proton- and neutron-target
observables are necessary to study $I=1/2$ transitions and isolate the separate $\gamma p N^*$ proton and $\gamma n N^*$ neutron photo-couplings~\citep{Sandorfi:2013cya}.
Information from neutron targets is comparatively scarce~\citep{ Anisovich:2017afs},  particularly in the hyperon channels~\citep{Compton_PhysRevC.96.065201,  AnefalosPereira:2009zw},
which is why the present measurement is of value.   Furthermore, the hyperon photoproduction channels   $\gamma N\rightarrow K \Lambda (\Sigma^{0})$
are  attractive for analysis for two reasons.  First,  the threshold for two-body hyperon final states is at $W \simeq 1.6$~GeV, above which lie numerous poorly known resonances.   Two-body strange decay modes, rather than cascading non-strange many-body decays, may be easier to interpret.
Second, the hyperon channels give easy access to recoil polarization observables on account of their self-analyzing weak decays.  While the present work  does not involve measurement of hyperon polarizations, previous work has shown the benefit of using such information to extract properties of higher-mass nucleon resonances~\citep{Paterson:2016vmc,Bradford_CxCz, Bradford_xsec, McNabb, Anisovich:2007bq, Paterson:2016vmc, McCracken, Dey, Anisovich:2017ygb}.  Thus, 
pursuing ``complete" amplitude information in the hyperon photoproduction channels can be complimentary to the analogous quest in, say, pion photoproduction.

In this article, we present first-time measurements of the beam-target observable $E$ on a longitudinally polarized neutron bound in deuterium in the quasi-free reaction $\gamma n(p) \to K^0Y(p)$.
The helicity asymmetry $E$  is formally defined as the normalized difference in photoproduction yield between  antiparallel ($\sigma^{A}$) and parallel ($\sigma^{P}$) configurations, {\it i.e.},  settings where the incident photon beam polarization is antialigned or aligned, respectively, with the longitudinal polarization of the target.  Following Ref.~\cite{Barker:1975bp} and Ref.~\cite{Sandorfi:2010uv}  write 
\begin{equation}
E=\frac{\sigma^{A}-\sigma^{P}}{\sigma^{A}+\sigma^{P}}.
\end{equation}
This helicity asymmetry, $E$, is related to the  cross section by
\begin{equation}
\left(\frac{d\sigma}{d\Omega}\right)=\left(\frac{d\sigma}{d\Omega}\right)_{0}\left(1-P_{T}P_{\odot}E\right),
\label{equation2}
\end{equation}
where $\left(d\sigma / d\Omega\right)_{0}$ is the differential cross section averaged over initial spin states and summed over the final states, and $P_{T}$
and $P_{\odot}$ are the target longitudinal and beam circular polarizations, respectively. 

The asymmetry results obtained are compared with several model predictions.  The first is a single-channel effective Lagrangian approach,  
KaonMAID~\citep{Mart:1999ed,Lee:1999kd}, with parameter constraints largely imposed from SU(6).  Without experimental constraints on the $N^* \Lambda K^0$ and  $\gamma n N^*$ vertices, the reaction of interest is difficult to model accurately.   
The second model giving predictions for the present results is the data description given by  SAID~\citep{SAID, Adelseck:1986fb}.   In general, SAID is more up to date than KaonMAID;  for the present reaction channels the SAID predictions are a polynomial fit to all available data  before  2008, assuming final state interactions for these polarization observables can be neglected~\cite{strakovsky}.
The third  comparison is made to the multichannel K-matrix formalism of the  Bonn-Gatchina~\citep{Anisovich:2012ct} group, which is the most up to date, being constrained by recent first-time measurements~\citep{Compton_PhysRevC.96.065201}  of the differential cross section for the reaction $\gamma n(p) \to K^0\Lambda (p)$  [with $(p)$ as the spectator proton].


\section{Experimental Procedures
\label{sec:Section-II}} 

The experiment was performed at the Thomas Jefferson National Accelerator Facility (JLab) using the CEBAF Large Acceptance
Spectrometer (CLAS)~\citep{CLAS-NIM}.  
This setup has been used for several studies of $K^+$ photoproduction of  
hyperonic final states on a proton target~\citep{Bradford_CxCz, Bradford_xsec, McNabb, Moriya:2014kpv, Moriya:2013hwg, Moriya:2013eb, McCracken, Dey} 
and on an effective neutron (deuteron) target~\citep{Compton_PhysRevC.96.065201,AnefalosPereira:2009zw}.
The present results stem from the so-called ``\textit{g14}" run period between December 2011 and May 2012, from which non-strange 
results have been previously reported~\citep{Ho:2017kca}.   
The CEBAF accelerator provided longitudinally polarized electron beams with energies of 
$E_{e}=2.281$,  
$2.257$,  
and $2.541$ GeV,  
and an \textit{average} electron beam polarization for the present study of $P_{e}=0.82\pm0.04$, which was measured routinely
by the Hall-B M\"oller polarimeter~\cite{Moller2}. The electron beam helicity was pseudorandomly flipped between +1 and $-1$ with a 960 Hz flip
rate. The electron beam was incident on the thin gold radiator of the Hall-B Tagger system~\citep{Sober} and produced circularly polarized
tagged photons. The polarization of the photons was determined using the Maximon and Olsen formula~\citep{Olsen:1959zz}
\begin{equation}
P_{\odot}=P_{e}\frac{4k-k^{2}}{4-4k+3k^{2}},
\end{equation}
where $P_{\odot}$ and $P_{e}$ are the photon and electron polarizations, respectively, and $k=E_{\gamma} / E_e$ is the ratio between
the photon energy and the electron beam energy. 

A 5-cm-long solid target of hydrogen deuteride (HD)  was used in the experiment~\citep{Lowry:2016uwa,Bass:2013noa}.
It  achieved vector polarizations of   25\%-30\% for deuterons, i.e., for 
{\it bound} neutrons in the deuteron with relaxation times of about a year.  
The polarized target was held at the center of CLAS using an in-beam cryostat that produced a 0.9~T holding field and operated at 50 mK.  The target polarization was monitored using nuclear magnetic resonance measurements~\citep{Lowry:2016uwa}.   The orientation of the target longitudinal polarization direction was inverted between periods of data taking, either parallel or antiparallel to the direction of the incoming photon beam.   Background events from the unpolarizable target wall material and aluminum cooling wires~\citep{Bass:2013noa} were removed using empty-target data, as discussed in Secs.~\ref{sec:Section-IIIa} and~\ref{sec:Section-IIIb}.

The specific reaction channel for this discussion came from events of the type $\gamma d \to \pi^+ \pi^- \pi^- p (X)$ using a readout trigger requiring a minimum of two charged particles in different CLAS sectors.  After particle identification  we required  the ``spectator,"  $X$, to be an undetected  low-momentum proton and possibly a photon, via the missing mass technique, as explained in the next section. In order to determine the $E$ asymmetry experimentally, the event yields in a given kinematic bin of $W$ and kaon center-of-mass angle were obtained by counting events with total c.m. helicity $h=$3/2 (laboratory-frame antiparallel configuration), called $N_{A}$, and events with $h=$1/2 (laboratory frame parallel configuration) called $N_{P}$, respectively. The $E$ observable 
was then computed as 
\begin{equation}
E=\frac{1}{\overline{P_{T}}\cdot\overline{P_{\odot}}}\left(\frac{N_{A}-N_{P}}{N_{A}+N_{P}}\right),
\label{Eq_Eval}
\end{equation}
where $\overline{P_{T}}$ and $\overline{P_{\odot}}$ are the run-averaged
target and beam polarizations, respectively.


\section{Data Analysis 
\label{sec:Section-III}}

The performance of the system was extensively studied for a reaction with much higher count rates than the present one.   The nonstrange reaction  $\gamma d \to \pi^- p (X) $ was investigated using many of the same analysis steps and methods discussed in this article to extract the $E$ observable for $\gamma n \to \pi^- p$~\citep{Ho:2017kca}.    The analysis steps outlined below were all tested on that reaction.   In particular, the boosted decision tree (BDT) selection procedure~\cite{DruckerCortes, ROE2005577} used below was validated against alternative ``cut-based" and kinematic fit  methods, with the result that the BDT procedure resulted in $\sim30\%$ larger yields of  signal events and therefore gave better statistical precision on the final $E$ asymmetry.

\subsection{Particle identification}
\label{sec:Section-IIIa}

For this particular analysis, we required that every selected event consists of at least two positive tracks and two negative tracks with associated photon tagger hits~\cite{Sober}.
The CLAS detector system determined the path length, the charge type,
the momentum and the flight time for each track~\citep{Sharabian,Mestayer,Smith}.
For each track of momentum $\overrightarrow{p}$, we compared the measured time of flight, $TOF_{m}$, to a hadron's expected time of
flight, $TOF_{h}$, for a pion and proton of identical momentum and path length. 
CLAS-standard cuts were placed on the difference between the measured and the expected time of flight, $\triangle TOF=TOF_{m}-TOF_{h}$. We selected events for which the two positively charged particles were the proton and $\pi^{+}$, and the two negatively charged were both  $\pi^{-}$. Well-established CLAS fiducial cuts were applied to select events with good spatial reconstruction. 

\begin{figure}
\vspace{-0.5cm}
\includegraphics[width=0.5\textwidth]{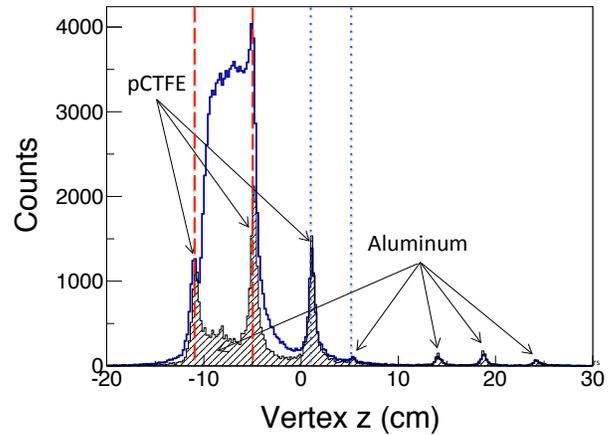}\protect
\vspace{-0.5cm}
\caption{The open histogram shows the vertex  distribution of events along the beam
line for a full target is the open histogram. Dashed red lines show the nominal target boundaries.  The peaks at $z>0$ are from target-independent foils in the cryostat;  the positions of two are  highlighted with dotted blue lines~\cite{Lowry:2016uwa}. The filled histogram shows the scaled target-empty  background  distribution.   
\label{Z-vertex} }
\end{figure}

Events originating from unpolarized target material\textit{\footnotesize{}\textemdash }aluminum cooling wires and polychlorotrifluoroethylene (pCTFE)\textit{\footnotesize{}\textemdash } dilute $E$ and must be taken into account. A  period of data taking was dedicated to an \textit{empty} target cell in
which the HD material was not present. This set of data was used to study and remove the bulk of the target material background on the basis of a loose missing mass cut.   
Figure \ref{Z-vertex} shows the resulting reconstructed reaction vertex for four-track data along the beam line both for a full target and for an empty target scaled to match the counts in several downstream target foils. 
The full-to-empty ratio of about 3.3:1 in the target region was important in selecting the optimal BDT cut discussed below.

\begin{figure}[htbp]
\vspace{-0.5cm}
\includegraphics[width=0.5\textwidth,height=0.35\textwidth]{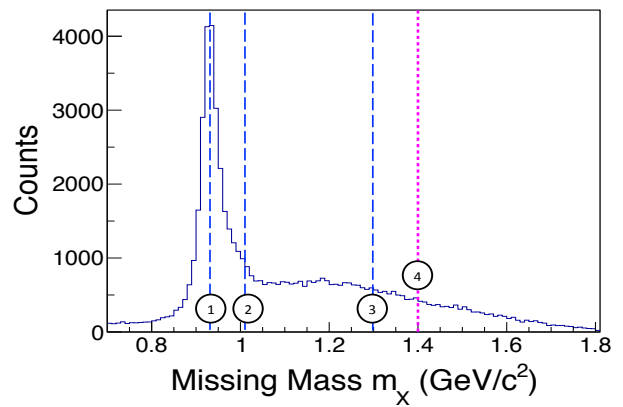}\protect
\vspace{-0.5cm}
\caption{The missing mass distribution, $\gamma d\rightarrow\pi^{-}\pi^{+}\pi^{-}pX$
after PID cuts showing the dominant spectator proton peak at ``1.''   The magenta line at ``4'' indicates a loose event rejection for $m_X >  1.4$~GeV/c$^2$. This rejects unambiguous
background but keeps $\Sigma^{0}\rightarrow\pi^{-}p(\gamma)$ events in which both a proton and a photon are missing between ``2'' and ``3.''  (See text.)
\label{MMass} }
\end{figure}

Figure \ref{MMass} shows the resulting target-full  missing mass distribution for spectator $X$ in $\gamma d\rightarrow\pi^{-}\pi^{+}\pi^{-}p(X)$,
after these cuts.  A clear peak corresponding to the spectator proton is seen at point 1 for events that produced a $\Lambda$ particle.  A loose cut was applied to reject events with missing mass larger than 1.4 GeV/c$^{2}$ at point 4 because of the presence of $\Sigma^{0}\rightarrow\pi^{-}p(\gamma)$ events. These have a 73-MeV photon in the final state in addition to the proton, and the distribution peaks at point 2 and has a kinematic tail to about point 3.

\subsection{$K^{0}Y$ event selection using BDT analysis}
\label{sec:Section-IIIb}

Because of the small reaction cross section in this experiment, a method was needed to optimally isolate the events of interest with minimal statistics loss.
The multivariate analysis tool called the boosted decision tree (BDT)  approach was used to select the exclusive events of interest in this study.   Three steps were needed to achieve this result.  The first BDT was created to select events from both the 
$\gamma d\rightarrow\pi^{-}\pi^{+}\pi^{-}p(p_{S})$ and the $\gamma d\rightarrow\pi^{-}\pi^{+}\pi^{-}p(p_{S}\gamma)$ final states, 
consistent with quasi-free production from a deuteron.  This was to reject target-material background and events with a high missing momentum of the undetected spectator nucleon, $p_S$.  The second BDT was created to remove the nonstrange pionic background with the same final states, that is, to pick out events with $\Lambda$ and  $\Sigma^0$ intermediate-state particles.    The third BDT was to separate the $K^{0}\Lambda$ and $K^{0}\varSigma^{0}$ events. 

This BDT algorithm is more efficient than a simple ``cut'' method in both rejecting background and keeping signal events~\citep{ROE2005577,Ho-thesis}.  The method  builds a ``forest'' of \textit{distinct} \textit{decision trees} that are linked together by a \textit{boosting} mechanism. Each decision tree constitutes a \textit{disjunction} of logical conjunctions (i.e., a graphical representation of a set of \textit{if-then-else} rules). Thus, the entire reaction phase-space is considered by every decision tree.  Before employing the BDT for signal and background classification, the BDT algorithm needs to be constructed (or trained) with \textit{training} data\textit{\textemdash }wherein  the category of every event is definitively known. We used the ROOT implementation of the BDT algorithm ~\citep{Hocker:2007zz}. 
Every event processed by the constructed BDT algorithm is assigned a value of between $-1$ and +1 that
quantifies how likely the processed event is a background event (closer to $-1$) or a signal event (closer to $+1$). An optimal cut on the
BDT output is chosen to maximize the\textit{ }$S/\sqrt{S+B}$ ratio, where $S$ and $B$ are the estimations, based on training data, of the initial number of signal
and background events, respectively.

The initial assignment of the $\pi^-$ particles to either $K^0$ or $\Lambda$ decay was studied with Monte Carlo simulation, and a loose selection based on invariant masses was made.  Specific details of these cuts are given in Ref.~\citep{Ho-thesis}. 

\begin{figure}[htbp]
\vspace{-0.5cm}
\includegraphics[width=0.5\textwidth]{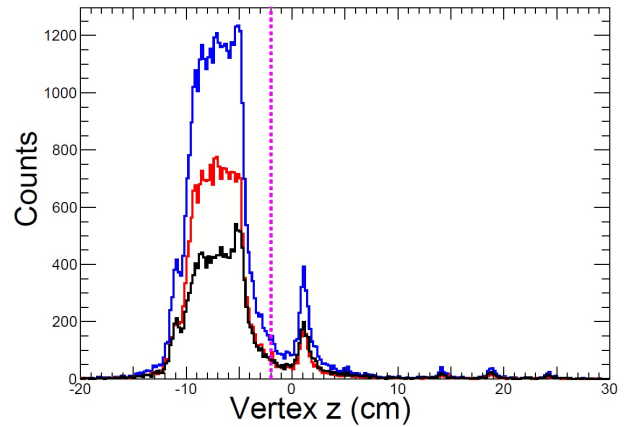}
\vspace{-1.cm}
\protect\caption{The reconstructed distribution of the reaction vertex along the beam
line showing target-full events in the top histogram (blue) after loose $K^0Y^0$ selection  and the missing mass cut shown in Fig.~\ref{MMass}.
Events selected by the first BDT are shown in the middle histogram (red), and rejected events in the bottom histogram  (black).
The magenta line indicates a loose cut to reject unambiguous target-material background. 
\label{Z_vertex_BDT} }
\end{figure}

The first BDT was trained using real empty-target data for the background training.   A signal Monte Carlo simulating quasifree hyperon production on the neutron was used for signal training data.  The momentum distribution of the spectator proton, $p_{s}$, followed the Hulth\`en potential~\citep{Cladis_PhysRev.87.425,Lamia:2012zz} for the deuteron. Based on this training, an optimal BDT cut  that maximized the estimated initial\textit{ }$S/\sqrt{S+B}$ ratio was selected. Figure \ref{Z_vertex_BDT} shows the total (blue histogram) and rejected (black histogram) events by the first BDT cut. In comparing  Figs.~\ref{Z-vertex} and \ref{Z_vertex_BDT}, two items should be noted. First, the BDT was trained to remove target-material background events with missing momentum not consistent with a Hulth\`en distribution. Second, the BDT background-rejection efficiency was not perfect, leaving some target-material background events that were removed in a subsequent step (Sec.~\ref{sec:Section-IIIc}). We then rejected events with $z>-2$~cm on the reaction vertex to remove remaining unambiguous background events due to various cryostat foils. 

\begin{figure}[htpb]
\includegraphics[width=0.48\textwidth]{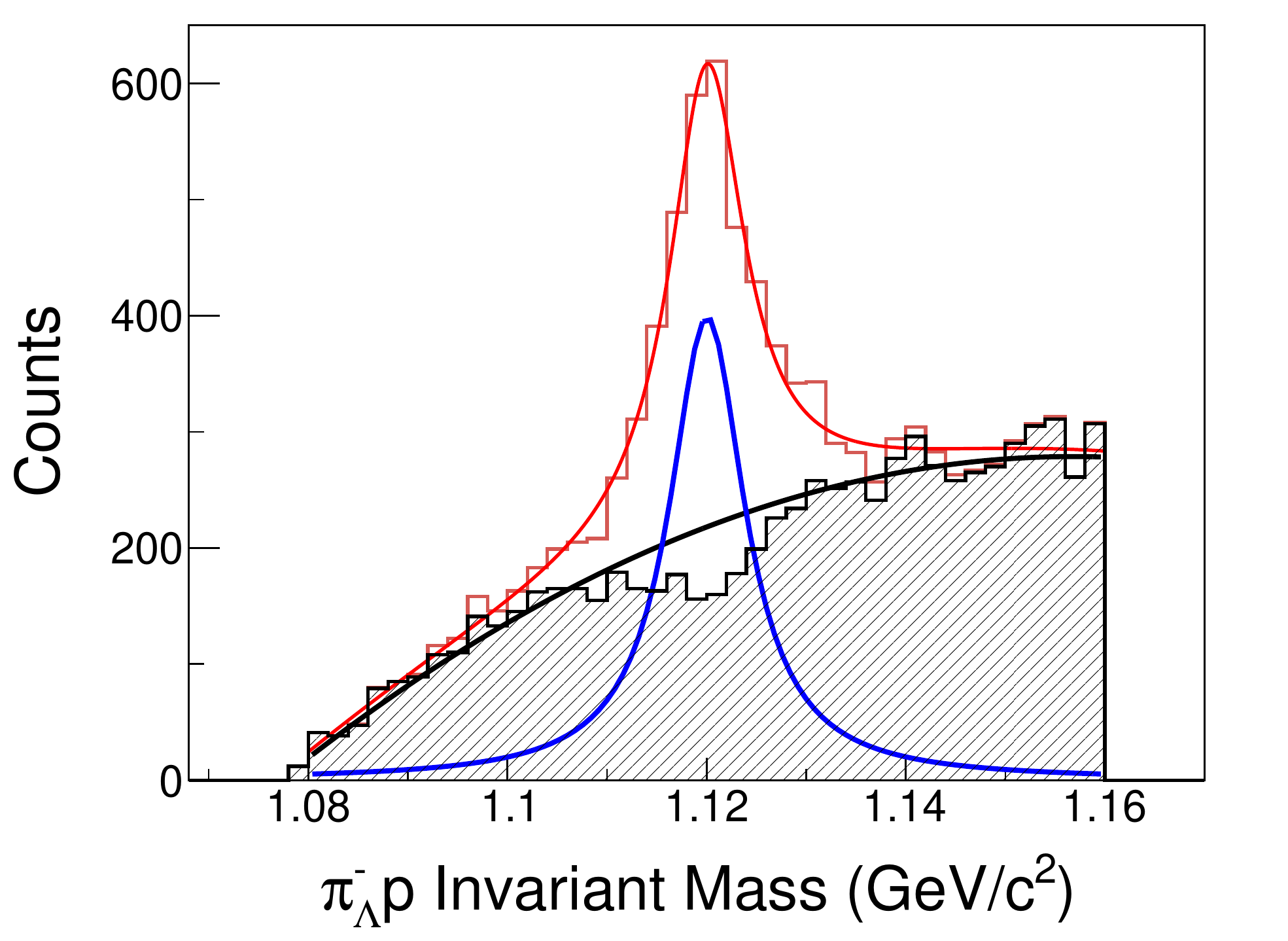}
\includegraphics[width=0.48\textwidth]{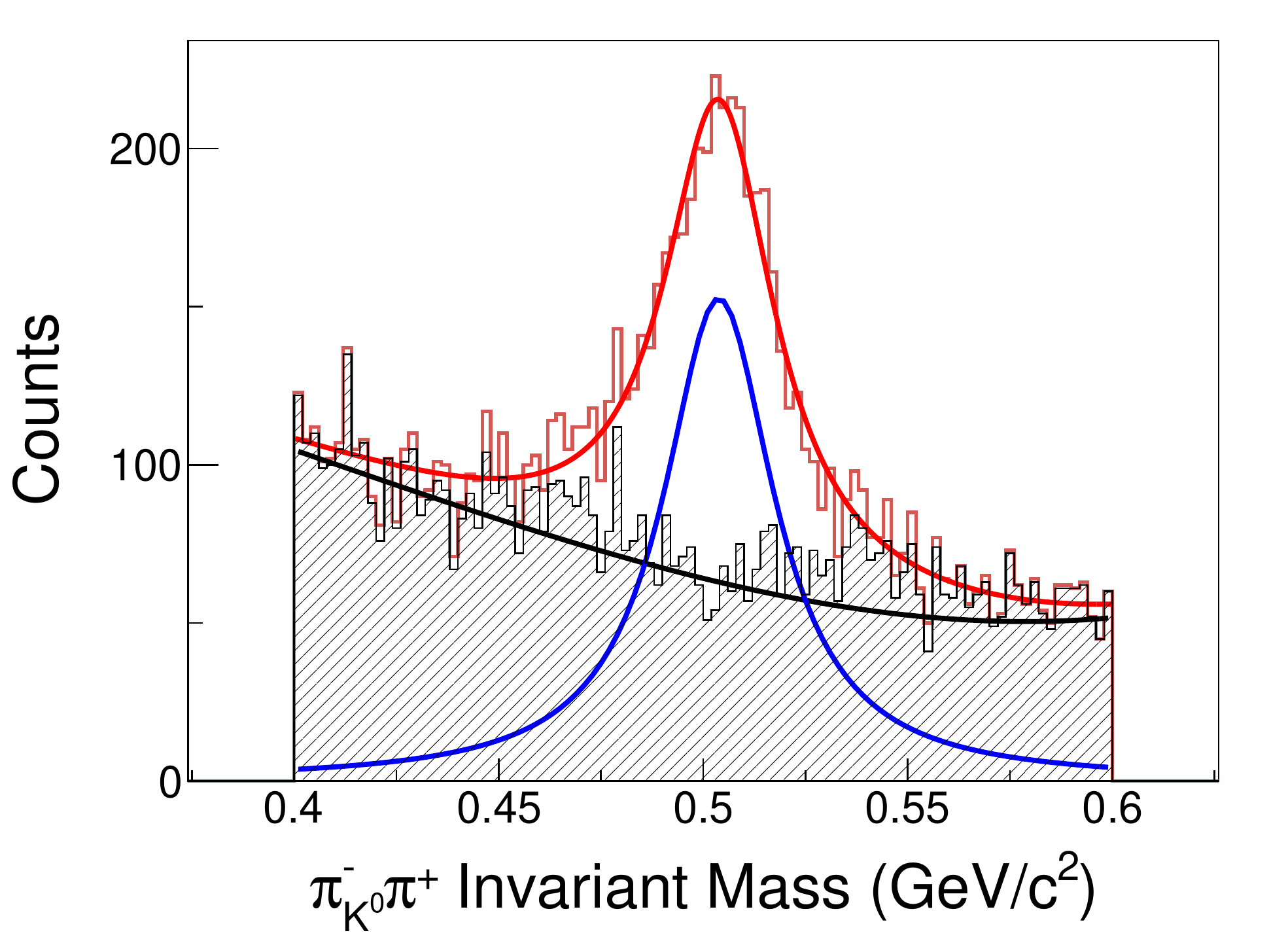}

\protect\caption{Invariant $\pi_{\Lambda}^{-}p$ mass (top) and invariant $\pi_{K^{0}}^{-}\pi^{+}$
mass (bottom) after target material background rejection by the first
BDT cut. Black histograms show events rejected  by the second BDT cut.  Fits of the sum (red curve) of a Breit-Wigner line-shape (blue curve) and a third order polynomial (black curve) are shown.  The fits aid the discussion in the text but were not used in the subsequent analysis. \label{lambda K0 IM} }
\end{figure}

The second-step BDT was trained using a four-body phase-space $\gamma d\rightarrow\pi^{-}\pi^{+}\pi^{-}p(p_{S})$
simulation as background training data and the $\gamma d\rightarrow K^{0}\Lambda(p_{S})$ simulation as signal training data. There were two negative pions in each event:  one from the decay of the $K^0$ and one from the decay of the hyperon.   The goal of the BDT analysis was to use the available correlations among all particles to sort the pions correctly and to select events with decaying strange particles.   The main training variables at this stage of the analysis included the 3-momenta of all the particles and the detached decay vertices of the $K^0$s and the hyperons.  
After the optimized BDT cut was placed, Fig.~\ref{lambda K0 IM}  shows the total (red histogram) and
rejected (black histogram) events after this second BDT analysis step.  The efficiency of the second BDT was less than 100\%, thus, there are remaining
target background events in the selected data sample.  The dips near the signal maxima in the background spectra show that the background is slightly undersubtracted. This issue is discussed and corrected below.  A fit with  a Breit-Wigner line shape and a polynomial was used to estimate  that the 
strange-to-non-strange ratio of events in the data set at this stage was about 2.3:1 in the peak regions.

\begin{figure*}[t]
\includegraphics[width=0.5\textwidth]{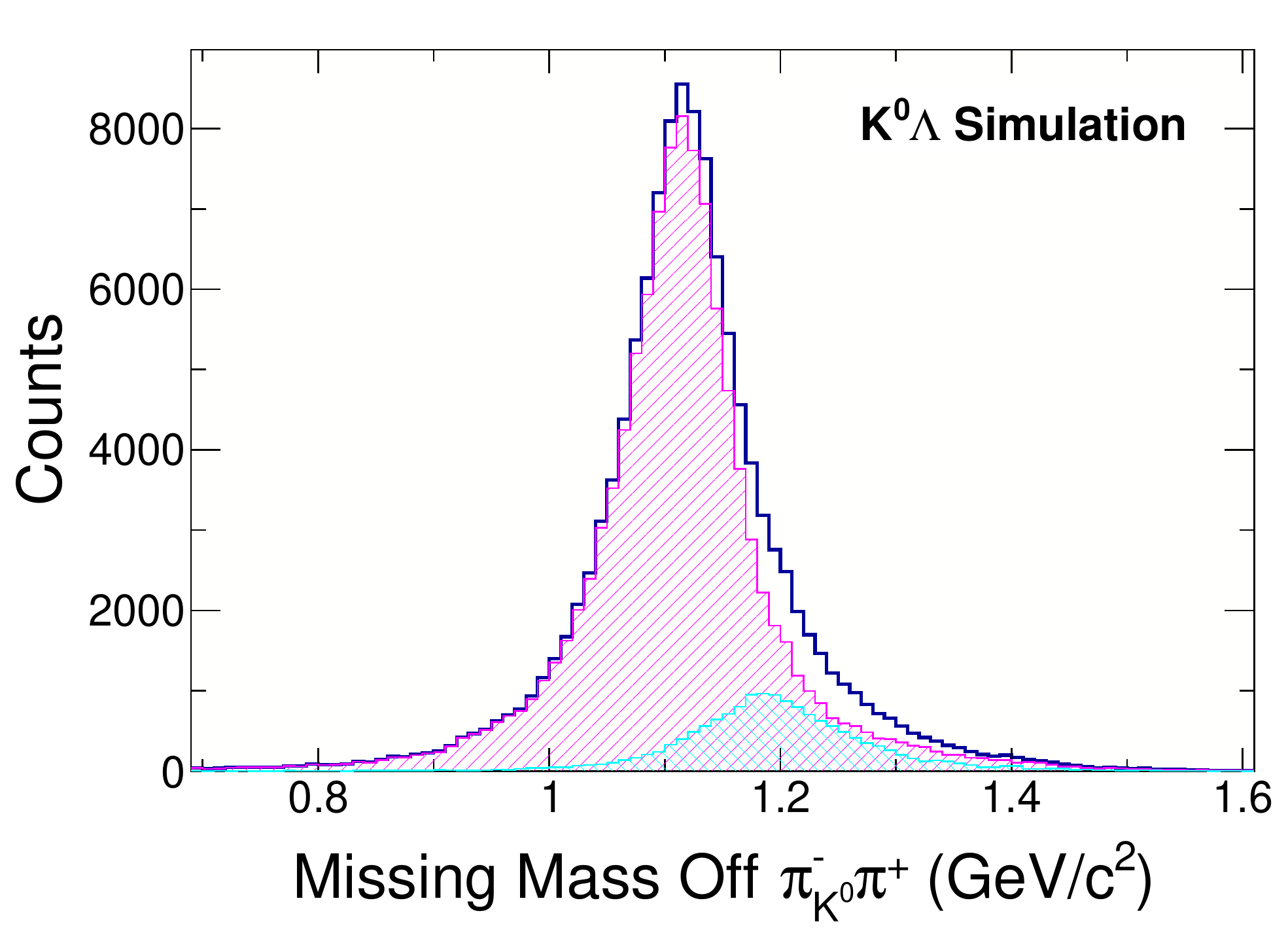}\includegraphics[width=0.5\textwidth]{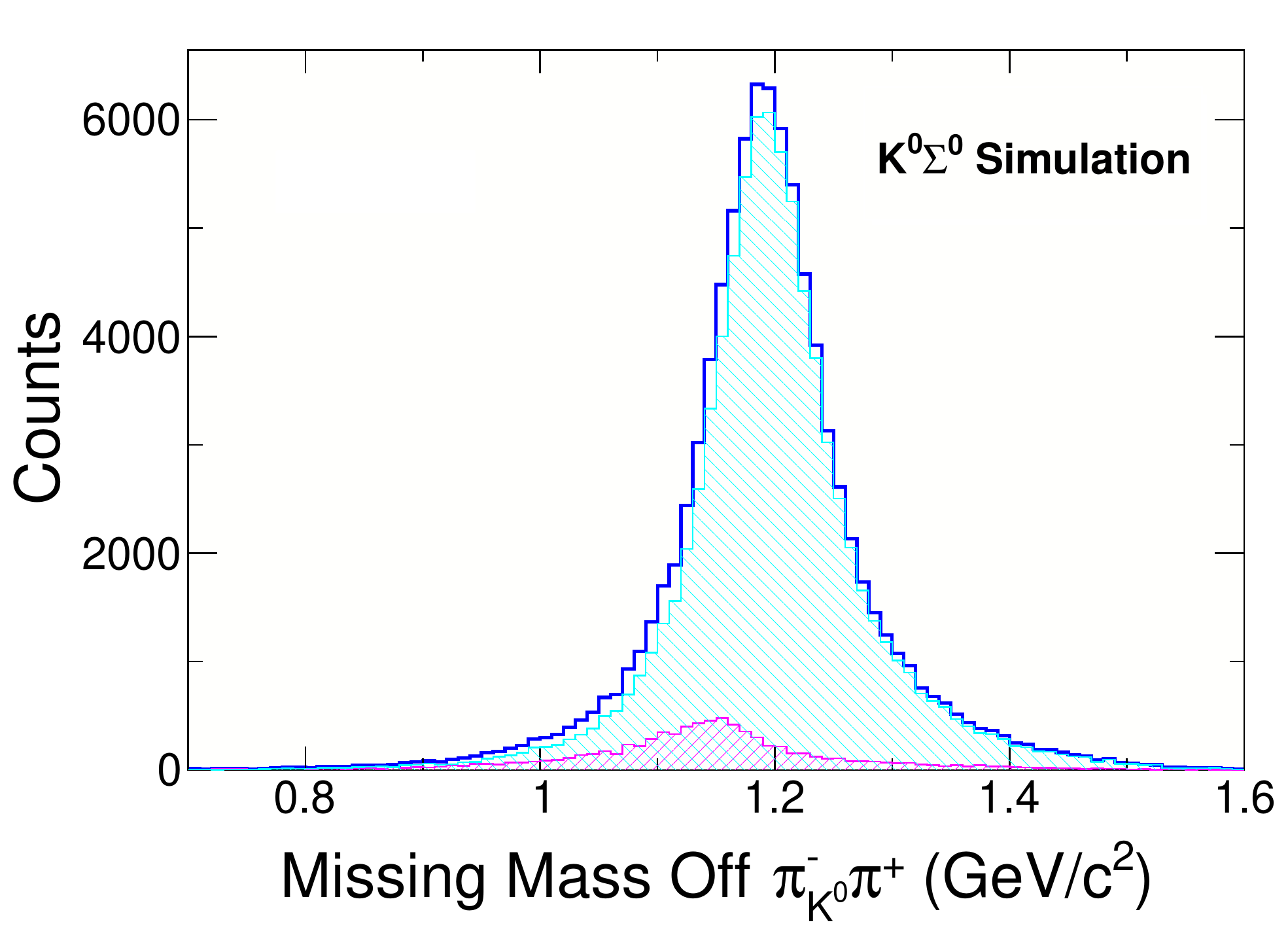}\protect
\caption{Distributions of missing mass from the reconstructed $K^{0}$, $\gamma n\rightarrow\pi_{K^{0}}^{-}\pi^{+}X$
for simulation data, assuming that the target is an at-rest neutron.
Left: the magenta histogram represents events with correct $K^{0}\Lambda$
classification, while the cyan histogram represents events with the wrong
$K^{0}\Sigma^{0}$classification. Right: the cyan histogram represents
events with the correct $K^{0}\Sigma^{0}$~classification, while the magenta
histogram represents events with the wrong $K^{0}\Lambda$ classification.
\label{MMass_off_K0_simulation} }
\end{figure*}

\begin{figure}[t]
\includegraphics[width=0.5\textwidth]{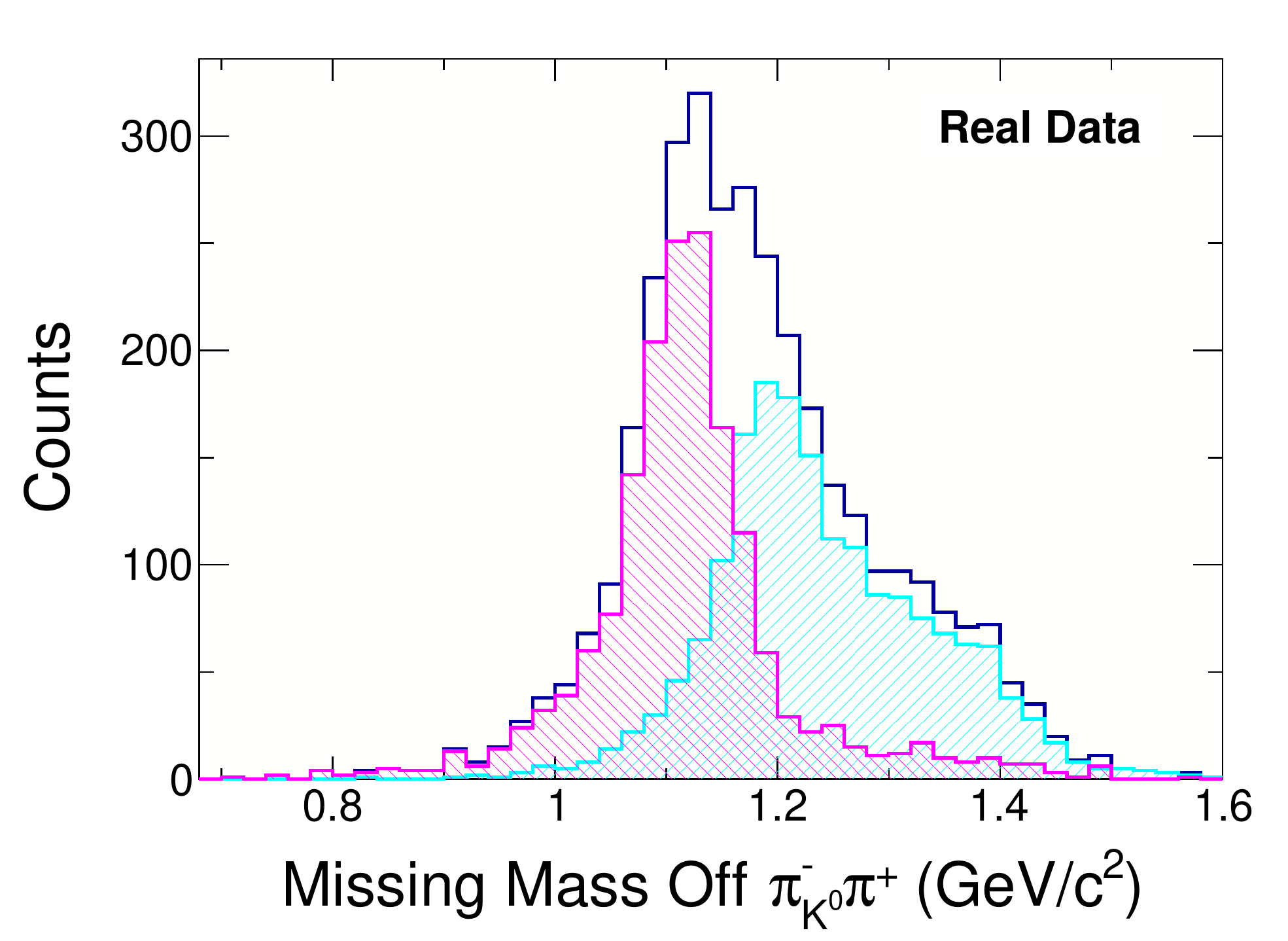}\protect\caption{Distribution of missing mass 
from the reconstructed $K^{0}$, $\gamma n\rightarrow\pi_{K^{0}}^{-}\pi^{+}X$
for real data, assuming that the target is an at-rest neutron,
after rejecting non-hyperon background by the second BDT cut.
The magenta (cyan) histogram was classified as $K^{0}\Lambda$ ($K^{0}\Sigma^{0}$)
using the third BDT selection step. \label{MMass_off_K0} }
\end{figure}

For the final task, separating the  $K^{0}\Lambda$ and $K^{0}\varSigma^{0}$ channels, the third BDT was trained using $\gamma d\rightarrow K^{0}\Sigma^{0}(p_{S})$ simulation as ``background'' training data and $\gamma d\rightarrow K^{0}\Lambda(p_{S})$ simulation as ``signal'' training data. Note that the term background used here is just for semantic convenience, since both channels were retained after applying the third optimized BDT cut.
Figure~\ref{MMass_off_K0_simulation}  shows in the left [right] histogram the classification success of the third BDT on $\gamma d\rightarrow K^{0}\Lambda(p_{S})$  [$\gamma d\rightarrow K^{0}\Sigma^{0}(p_{S})$] simulation data. The histograms reveal that a small number of $K^{0}\Lambda$ events  would be misclassified as $K^{0}\varSigma^{0}$ events, and vice versa. In the next section, the correction for the contamination on both final
data sets is discussed. Figure \ref{MMass_off_K0} shows the separation result from the third BDT on real data. 

\subsection{Corrections for remaining backgrounds and asymmetry calculation}
\label{sec:Section-IIIc}

The $E$ asymmetry values  for both target-material and non-strange background events were statistically
consistent with 0~\citep{Ho-thesis}; therefore, we implemented an  approximation procedure
to correct for the dilution effect from the remaining background.
We estimated two ratios: one for the remaining fraction of target background (TGT), $R^{TGT}$,  
and one for the fraction of remaining nonstrange (NS) final-state events mixed with the hyperon events, $R^{NS}$.
We write 
$R^{TGT}= {N^{remain}} / {N^{HD}}$,
and 
$R^{NS}={Y^{remain}}/{Y^{K^{0}Y}}$.   
$N^{remain}$ and $N^{HD}$ are the estimated number of remaining target-material background events
and true deuteron events after the first BDT and $z=-2$~cm vertex cuts, respectively.
$Y^{remain}$ and  $Y^{K^{0}Y}$ are the estimated number of remaining nonstrange  and true $K^{0}Y$ events after the second
BDT cut, respectively. Next, let $Y_{BDT}$  be the number of events that passed the $z$-vertex cut and the first two BDT selections; then $Y_{BDT}$
can be partitioned into 
\begin{align}
Y_{BDT}&=\left(1+R^{NS}\right)Y^{K^{0}Y} \nonumber \\
&=\left(1+R^{NS}\right)\left[Y_{HD}^{K^{0}Y}+Y_{TGT}^{K^{0}Y}\right],\;\label{eq:wideeq-2-3}
\end{align}
since $Y^{K^{0}Y}$ also comprises events from the remaining target-material
background and the bound signal events.  If we further allow  
$Y_{TGT}^{K^0 Y} / Y_{HD}^{K^0 Y} = N^{remain} / N^{HD} = R^{TGT}$,
then $Y_{BDT}$ can finally be expressed as
\begin{equation}
Y_{BDT}=\left(1+R^{NS}\right)\left(1+R^{TGT}\right)Y_{HD}^{K^{0}Y},\;\label{eq:wideeq-2}
\end{equation}
or
\begin{equation}
Y_{HD}^{K^{0}Y}=\left(1+R^{NS}\right)^{-1}\left(1+R^{TGT}\right)^{-1}Y_{BDT}.
\label{eq:wideeq-2-1}
\end{equation}
These relations should remain valid for both $Y_{BDT}^{K^{0}\Lambda}$ and $Y_{BDT}^{K^{0}\Sigma^{0}}$, 
which are the $K^{0}\Lambda$ and $K^{0}\Sigma^{0}$ signal events from bound neutrons, respectively.  
The backgrounds that leak through the BDT filters will be helicity independent and will be subtracted in the numerator of Eq.~(\ref{Eq_Eval}).  
Using Eq.~(\ref{eq:wideeq-2-1}) to correct the summed yields in the denominator gives the corrected asymmetry as
\begin{equation}
E_{corrected}^{K^{0}Y}=\left(1+R^{NS}\right)
\times\left(1+R^{TGT}\right)E_{BDT}^{K^{0}Y},\;\label{eq:wideeq-2-2}
\end{equation}
where $E_{BDT}^{K^{0}Y}$ is obtained from $Y_{BDT}^{K^{0}Y}$ (or,
more exactly, $Y_{BDT}^{P}$ and $Y_{BDT}^{A}$
of the $K^{0}Y$ parallel and antiparallel subsets). 
From the simulations we found average values of $R^{TGT}$ and $R^{NS}$ of 0.09  and 0.17, respectively, with some dependence on
the specific run period.

Next we discuss a correction for the third BDT classification result.
Recall that the third BDT selection separates the true signal $K^{0}Y$ events into two subsets: one is mostly $K^{0}\Lambda$ events,
and the other is mostly $K^{0}\varSigma^{0}$. If we denote  $N_{\Lambda}^{BDT}$ and $N_{\Sigma^{0}}^{BDT}$ as the
number of events the third BDT identified as $K^{0}\Lambda$ and $K^{0}\varSigma^{0}$ events, respectively, then we have the expressions

\begin{equation}
N_{\Lambda}^{BDT}=\omega_{\Lambda}N_{\Lambda}^{true}+(1-\omega_{\Sigma^{0}})N_{\Sigma^{0}}^{true},\;\label{eq:wideeq-3}
\end{equation}
\begin{equation}
N_{\Sigma^{0}}^{BDT}=(1-\omega_{\Lambda})N_{\Lambda}^{true}+\omega_{\Sigma^{0}}N_{\Sigma^{0}}^{true},\;\label{eq:wideeq-3-1}
\end{equation}
where $\omega_{\Lambda}$ and $\omega_{\Sigma^{0}}$ are the fractions of events correctly identified: these values were
estimated based on simulation data. After rearrangement, we arrive at the expressions 
\begin{align}
N_{\Lambda}^{true}&=\left[\omega_{\Lambda}-\frac{(1-\omega_{\Sigma^{0}})}{\omega_{\Sigma^{0}}}(1-\omega_{\Lambda})\right]^{-1} \nonumber \\
&\times \left[N_{\Lambda}^{BDT}-\frac{(1-\omega_{\Sigma^{0}})}{\omega_{\Sigma^{0}}}N_{\Sigma^{0}}^{BDT}\right],\;
\label{eq:wideeq}
\end{align}
\begin{align}
N_{\Sigma^{0}}^{true}&=\left[\omega_{\Sigma^{0}}-\frac{(1-\omega_{\Lambda})}{\omega_{\Lambda}}(1-\omega_{\Sigma^{0}})\right]^{-1} \nonumber \\ 
&\times \left[N_{\Sigma^{0}}^{BDT}-\frac{(1-\omega_{\Lambda})}{\omega_{\Lambda}}N_{\Lambda}^{BDT}\right]\;
\label{eq:wideeq-1}.
\end{align}
%

The \textit{corrected} $E$ asymmetry was obtained using the derived $N_{\Lambda}^{true}$ and $N_{\Sigma^{0}}^{true}$ by using
Eq.~(\ref{Eq_Eval}).  From the simulations we found average values of $\omega_Y$ of 0.87 and 0.91 for $\Lambda$ and $\Sigma^0$ events, respectively.

The neutron polarization in the deuteron is smaller than the deuteron polarization because the deuteron wavefunction has, in addition to an $S$-wave component, a $D$-wave component in which the spin of the neutron need not be aligned with the deuteron spin.   This was studied using data for the $\gamma n \to \pi^- p $ reaction and reported in our previous publication~\citep{Ho:2017kca}.  It was found that for spectator recoil momenta of less than 100~MeV/$c$ the correction was negligible.  Had we cut on the recoil momentum at 200~MeV/$c$ rather than 100~MeV/$c$, a measured dilution factor of $(8.6\pm0.1)$\%  would have been necessary for the nonstrange channel.   But different reaction channels  may exhibit different sensitivities to recoil momentum.  For the reaction under discussion here we could not afford the statistical loss by cutting on recoil momentum, and we elected to make a conservative correction based on the general considerations in \citep{Ramachandran:1979ck}.  The neutron polarization can
be estimated as $P_{n}=P_{d}(1-\frac{3}{2}P_{D})$, where $P_{n}$ and $P_{d}$ are neutron and deuteron polarizations, respectively, and $P_{D}$ denotes the deuteron $D$-state probability.  The latter is not strictly an observable and needs only to be treated consistently within a given $NN$ potential.
Following  Ref.~\citep{Ramachandran:1979ck}, we take the $D$-state contribution averaged over a range of $NN$ potentials as about  5\%, which implies that the neutron polarization is  92.5\% of the deuteron polarization, or a 7.5\% dilution factor. 

\subsection{Systematic Uncertainties}

We implemented four systematic studies to quantify the robustness of the trained BDT algorithms and the sensitivity of our results on
the  correction procedures introduced in the previous section.  
Two  tests studied the effect of loosening the first and the second BDT cuts, respectively.  One test focused on the sensitivity of the $E$ results on the third correction\textit{\textemdash }the correction procedure that was implemented to ``purify'' the final selected $K^{0}\Sigma^{0}$($K^{0}\Lambda$) sample.  Finally, we reduced the beam and target polarizations by one standard deviation of their respective total uncertainties (statistical and systematic) to study the changes in the $E$ results.

Finally, we note a complication that could occur when summing $\Lambda$ yields to create the $E$ asymmetries. The relative angular distribution between the $\pi^-$ and the $p$ that are used to reconstruct a $\Lambda$ carries information on the recoil polarization of the latter. When summed over azimuthal angles, this information is lost. However, limitations in detector acceptance could result in  incomplete integration, which in principle could introduce into Eq.~\ref{equation2}  a dependence on six additional observables~\cite{Sandorfi:2010uv}. The gaps in CLAS acceptance are modest, and due to the lower than expected production cross sections, the data below are presented in broad kinematic bins, which tends to dilute such effects. On the scale of our statistical uncertainties, such corrections are expected to be negligible and we have not attempted to correct for them.


\section{Results \label{sec:Section-IV}}

\begin{figure}[htbp]
\vspace{-0.5cm}
\includegraphics[angle=-90 , width=0.55\textwidth, trim=0 3.5cm 0 0, clip]{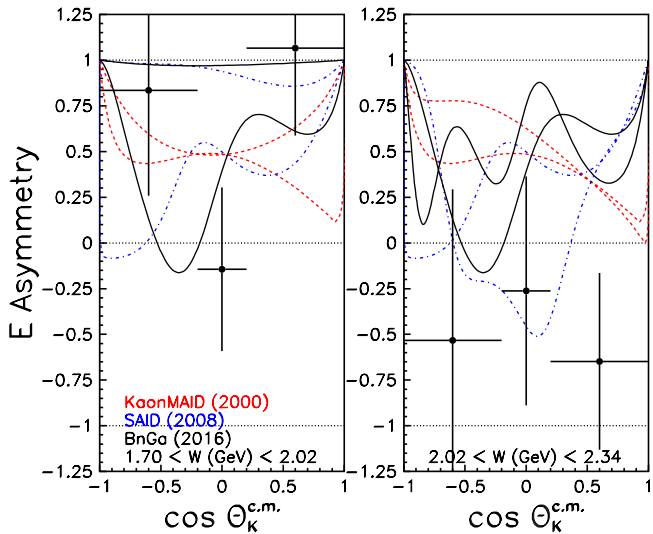}
\vspace{-10mm} 
\caption{Helicity asymmetry  $E$ for the  ${K^{0}\Lambda}$ final state (with combined statistical and systematic uncertainties) vs. $\cos\theta_{K^{0}}$   The asymmetries are shown with the neutron-target theoretical models KaonMaid ~\citep{Mart:1999ed} (dashed red curve) and SAID~\citep{SAID} (dot-dashed blue curve) and Bonn-Gatchina~\citep{Anisovich:2007bq,Anisovich:2012ct} (solid black curve).  Because of the 0.32-GeV-wide $W$ bins, each model is represented by two curves, computed at the bin endpoint $W$ values, as labeled. 
\label{Ecostheta_K0L} }
\end{figure}

We present here the results for the $E$ asymmetry in two $W$ energy bins. The lower bin is from 1.70 to 2.02 GeV and denoted $W_{1}$, while the higher bin is from 2.02 to 2.34 GeV and referred to  $W_{2}$.  Due to small cross sections for $K^0Y$ photoproduction, and to detector inefficiencies that are amplified by the  required identification of four charged particles, our statistics are  sufficient for only three bins in the $K^0$ center-of-mass production angle.  The measurements for the $\gamma n \to K^0 \Lambda$ reaction  are plotted together with predictions from the KaonMAID, SAID, and Bonn-Gatchina (BnGa)  models in Fig.~\ref{Ecostheta_K0L}.   The data show that the $K^0\Lambda$ asymmetry is largely positive below 2~GeV and mostly negative above 2~GeV, without more discernible trends.   Values of $E$ must approach $+1$ at $\cos \theta^{c.m.}_{K^0}\to \pm 1$ to conserve angular momentum.   Thus, the values for $E$  in bin $W_2$ must change rather rapidly near the extreme angles.

For comparison, the PWA combine results from many experiments at different energies, and this results in varying degrees of sensitivity to energy and angle. This is illustrated in Fig.~\ref{Ecostheta_K0L} by the SAID and BnGa PWA predictions at the limits of the energy bins. None of the models were tuned to these results;  that is, the models are all predictions based on  fits to previously published data on other observables.
First, one observes that the data are not statistically strong enough to strongly discriminate among the models.  In the lower $W$ bin all three models
can be said to agree with the data.   In the higher $W$ bin the SAID model may be  slightly favored by the data among the three.

\begin{figure}[htbp]
\vspace{-0.5cm}
\includegraphics[angle=-90 ,  width=0.55\textwidth, trim=0 3.5cm 0 0, clip]{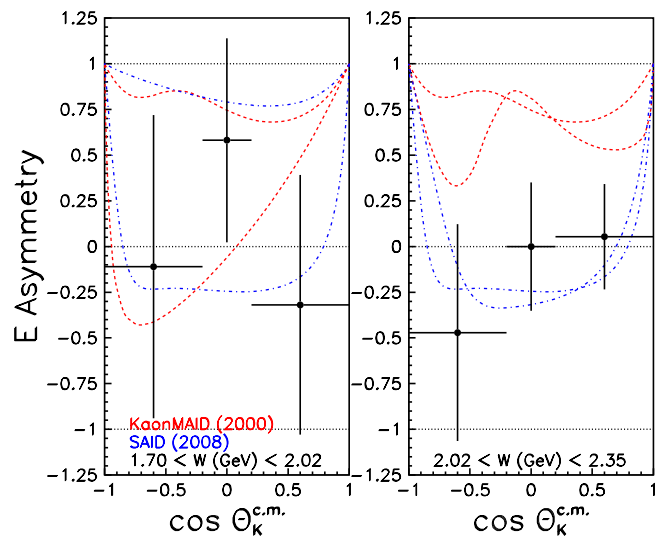}
\vspace{-10mm} 
\caption{Helicity asymmetry  $E$ for the  ${K^{0}\Sigma^0}$ final state (with combined statistical and systematic uncertainties) vs. $\cos\theta_{K^{0}}$ for two 0.32-GeV-wide  energy bands in $W$, as labeled. Model curves are as in Fig.~\ref{Ecostheta_K0L}.
\label{Ecostheta_K0S} }
\end{figure}

The results for the $\gamma n \to K^0 \Sigma^0$ channel are plotted in Fig.~\ref{Ecostheta_K0S}, together with model predictions from SAID and Kaon-MAID.   In contrast to the $K^0 \Lambda$  channel at lower $W$, here the data hint at less positive values for $E$.   
In the bin for  $W$ above 2 GeV, the data are also consistent with 0 for $K^0\Sigma^0$, whereas the $K^0\Lambda$ data tended to be negative.  In fact, the $K^0\Sigma^0$ asymmetry is consistent with 0 in all available bins.   
The model comparisons show that the KaonMAID prediction for the $K^0\Sigma^0$ channel in the higher $W$ bin are probably not consistent with the data, while the SAID result is consistent with the data.   For the $K^0\Sigma^0$  case we do not have predictions from the Bonn-Gatchina model  because the unpolarized differential cross section has not been measured yet, and without it the model  does not have a prediction available. 

\begin{figure}[htbp]
\vspace{-0.5cm}
\includegraphics[angle=-90 ,  width=0.55\textwidth, trim=0 3.5cm 0 0, clip]{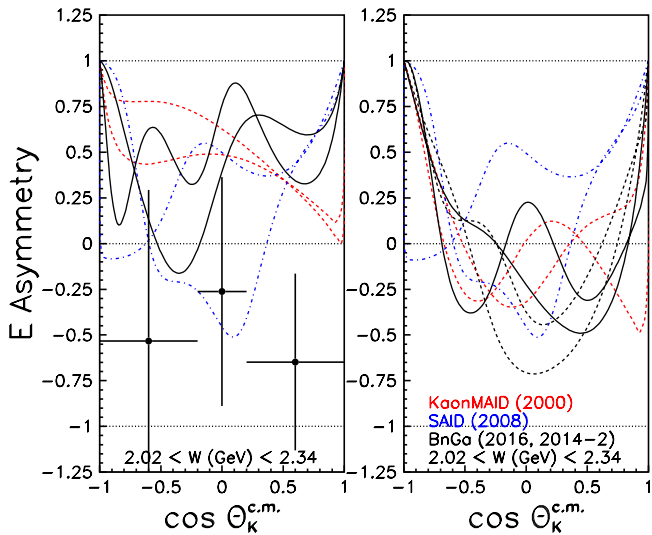}
\vspace{-10mm} 
\caption{
Helicity asymmetry  $E$ for the  ${K \Lambda}$ final state vs. $\cos\theta_{K^{0}}$ for energy band  $W_2$.   Left: Data from Fig.~\ref{Ecostheta_K0L} together with model  predictions for a neutron target. Right: Model calculations  for the $K^+ \Lambda$ reaction on a proton target, as computed using KaonMaid~\citep{Mart:1999ed} (dashed red curve),  SAID~\citep{SAID} (dot-dashed blue curve) and Bonn-Gatchina~\citep{Anisovich:2007bq,Anisovich:2012ct} (solid and dashed black curves).   Curves on the right are closer to the (reaction mismatched) data shown on the left.
\label{Ecostheta_K0L2} }
\end{figure}

%
In order to show  one other comparison between data and theory, we plot some of the present results for a neutron target together with the model predictions for the $K^+ \Lambda$ reaction on a {\textit {proton}} target in Fig.~\ref{Ecostheta_K0L2}.  This is intended to show the difference in  the model predictions on protons versus neutrons.  One sees how different the three model predictions are for protons versus neutrons.   One notes that the predictions for the proton target calculations all tend to be closer to the new data we are presenting for a neutron target.   This suggests that calculations of the $E$ observable for a neutron target can be improved.   
Thus, we may expect these present results to have some impact on the further development of these models.

So far unpublished CLAS results for the corresponding reaction $\gamma p \to K^+ \Lambda$ have higher statistics and finer energy bins than the present results (since the identification of this final state requires the detection of fewer particles).   The present  $K^0 \Lambda$ results are, within our uncertainties, similar to the $K^+\Lambda$ asymmetries in Ref.~\cite{LiamCasey}. The numerical values of the measured  $K^0 \Lambda$  and $K^0 \Sigma^0$ $E$ asymmetries, together with their statistical and systematic uncertainties, are reported in Table \ref{Tab:E_sys_stat}.

\begin{table*}[htbp]
\hfill{}%
\begin{tabular}{ccccc}
\hline\hline 
\multicolumn{1}{c}{   } &  & \multicolumn{3}{c}{$\cos\theta_{K^{0}}$}\tabularnewline
\cline{3-5} 
\multicolumn{1}{c}{} &  & $-$0.6 & 0.0 & $+$0.6\tabularnewline
\hline 
\multirow{2}{*}{$K^{0}\Lambda$ } & $W_{1}$ & 0.834$\pm$0.499$\pm$0.287 & $-$0.144$\pm$0.436$\pm$0.098 & 1.066$\pm$0.419$\pm$0.231\tabularnewline
\cline{2-5} 
 & $W_{2}$ & $-$0.533$\pm$0.752$\pm$0.345 & $-$0.263$\pm$0.618$\pm$0.101 & $-$0.648$\pm$0.464$\pm$0.136\tabularnewline
\hline 
\multirow{2}{*}{$K^{0}\Sigma^{0}$} & $W_{1}$ & $-$0.110$\pm$0.723$\pm$0.406 & 0.581$\pm$0.539$\pm$0.144 & $-$0.319$\pm$0.541$\pm$0.460\tabularnewline
\cline{2-5} 
 & $W_{2}$ & $-$0.471$\pm$0.446$\pm$0.391 & 0.0002$\pm$0.317$\pm$0.150 & 0.054$\pm$0.281$\pm$0.065\tabularnewline
\hline \hline
\end{tabular}
\hfill{}

\protect\caption{Numerical values of the $E$ asymmetry measurements for the $K^{0}\Lambda$/$K^{0}\Sigma^{0}$
channels. The uncertainties are statistical and systematic, respectively.  Center-of-mass energy ranges are $1.70 < W_1 < 2.02$~GeV and $2.02 < W_2 < 2.34$~GeV.
\label{Tab:E_sys_stat}}
\end{table*}

\section{Conclusions \label{sec:Section-V}}

We have reported the first set of $E$ asymmetry measurements for the reaction $\gamma d\rightarrow K^{0}Y(p_{s})$ for 1.70~GeV$\leq W \leq$ 2.34~GeV. In particular, we have described the three-step BDT-based analysis method developed to select a clean sample of $p\pi^{+}\pi^{-}\pi^{-}$ with intermediate hyperons. We have plotted the $E$ asymmetry as a function of $\cos \theta_{K^{0}}^{CM}$.
Several systematic uncertainty tests led to the conclusion that statistical uncertainties dominated the final results.  The numerical values of the measured $E$ asymmetries and their statistical and systematic uncertainties are reported in Table \ref{Tab:E_sys_stat}.

Evidently,  this analysis is limited by the small cross sections of the channels of interest, leading to large uncertainties in the measurements of the $E$ asymmetry.  At present, comparison with several models makes no decisive selections among the model approaches.  
Overall, the BnGa predictions are of a quality similar to that of the SAID predictions. The Kaon-MAID predictions for both channels seem less successful.  Among all three model comparisons, the distinction between proton- and neutron-target predictions are  differentiated by the data:  The proton-target predictions compare better than the neutron-target predictions with the experimental results.  In principle, this information is valuable since it hints at the necessary isospin decomposition of the hyperon photoproduction mechanism.   

At present, multipole analyses for $K^0Y$ channels are severely limited by the available data.   Higher statistical data on these channels for a number of other polarization  observables, from a much longer (unpolarized) target, have been collected during the $g13$ running period with CLAS and are under analysis.   A greater number of different polarization observables is generally more effective than precision at determining the  photoproduction amplitude~\cite{Sandorfi:2010uv}.   When these $g13$ results become available, the present data on the beam-target $E$ asymmetry are likely to have a larger impact.   

\begin{acknowledgments}
We acknowledge the outstanding efforts of the staff of the Accelerator
and Physics Divisions at Jefferson Lab who made this experiment
possible. The work of the Medium Energy Physics group at Carnegie
Mellon University was supported by DOE Grant No. DE-FG02-87ER40315.  The
Southeastern Universities Research Association (SURA) operated the
Thomas Jefferson National Accelerator Facility for the United States
Department of Energy under Contract No. DE-AC05-84ER40150.  Further
support was provided by 
the National Science Foundation, 
the Chilean Comisi\'on Nacional de Investigaci\'on Cient\'ifica y Tecnol\'ogica (CONICYT),
the French Centre National de la Recherche Scientifique,
the French Commissariat \`{a} l'Energie Atomique,
the Italian Istituto Nazionale di Fisica Nucleare,
the National Research Foundation of Korea,
the Scottish Universities Physics Alliance (SUPA),
and the United Kingdom's Science and Technology Facilities Council.
\end{acknowledgments}


\bibliography{CLAS_K0Y_E_photoproduction}

\begin{thebibliography}{62}%
\makeatletter
\providecommand \@ifxundefined [1]{%
 \@ifx{#1\undefined}
}%
\providecommand \@ifnum [1]{%
 \ifnum #1\expandafter \@firstoftwo
 \else \expandafter \@secondoftwo
 \fi
}%
\providecommand \@ifx [1]{%
 \ifx #1\expandafter \@firstoftwo
 \else \expandafter \@secondoftwo
 \fi
}%
\providecommand \natexlab [1]{#1}%
\providecommand \enquote  [1]{``#1''}%
\providecommand \bibnamefont  [1]{#1}%
\providecommand \bibfnamefont [1]{#1}%
\providecommand \citenamefont [1]{#1}%
\providecommand \href@noop [0]{\@secondoftwo}%
\providecommand \href [0]{\begingroup \@sanitize@url \@href}%
\providecommand \@href[1]{\@@startlink{#1}\@@href}%
\providecommand \@@href[1]{\endgroup#1\@@endlink}%
\providecommand \@sanitize@url [0]{\catcode `\\12\catcode `\$12\catcode
  `\&12\catcode `\#12\catcode `\^12\catcode `\_12\catcode `\%12\relax}%
\providecommand \@@startlink[1]{}%
\providecommand \@@endlink[0]{}%
\providecommand \url  [0]{\begingroup\@sanitize@url \@url }%
\providecommand \@url [1]{\endgroup\@href {#1}{\urlprefix }}%
\providecommand \urlprefix  [0]{URL }%
\providecommand \Eprint [0]{\href }%
\providecommand \doibase [0]{http://dx.doi.org/}%
\providecommand \selectlanguage [0]{\@gobble}%
\providecommand \bibinfo  [0]{\@secondoftwo}%
\providecommand \bibfield  [0]{\@secondoftwo}%
\providecommand \translation [1]{[#1]}%
\providecommand \BibitemOpen [0]{}%
\providecommand \bibitemStop [0]{}%
\providecommand \bibitemNoStop [0]{.\EOS\space}%
\providecommand \EOS [0]{\spacefactor3000\relax}%
\providecommand \BibitemShut  [1]{\csname bibitem#1\endcsname}%
\let\auto@bib@innerbib\@empty
\bibitem [{\citenamefont {Gross}\ and\ \citenamefont
  {Wilczek}(1973)}]{Gross:1973id}%
  \BibitemOpen
  \bibfield  {author} {\bibinfo {author} {\bibfnamefont {D.~J.}\ \bibnamefont
  {Gross}}\ and\ \bibinfo {author} {\bibfnamefont {F.}~\bibnamefont
  {Wilczek}},\ }\bibfield  {title} {\enquote {\bibinfo {title} {{Ultraviolet
  Behavior of Nonabelian Gauge Theories}},}\ }\href {\doibase
  10.1103/PhysRevLett.30.1343} {\bibfield  {journal} {\bibinfo  {journal}
  {Phys. Rev. Lett.}\ }\textbf {\bibinfo {volume} {30}},\ \bibinfo {pages}
  {1343--1346} (\bibinfo {year} {1973})}\BibitemShut {NoStop}%
\bibitem [{\citenamefont {Politzer}(1973)}]{Politzer:1973fx}%
  \BibitemOpen
  \bibfield  {author} {\bibinfo {author} {\bibfnamefont {H.~D.}\ \bibnamefont
  {Politzer}},\ }\bibfield  {title} {\enquote {\bibinfo {title} {{Reliable
  Perturbative Results for Strong Interactions?}}}\ }\href {\doibase
  10.1103/PhysRevLett.30.1346} {\bibfield  {journal} {\bibinfo  {journal}
  {Phys. Rev. Lett.}\ }\textbf {\bibinfo {volume} {30}},\ \bibinfo {pages}
  {1346--1349} (\bibinfo {year} {1973})}\BibitemShut {NoStop}%
\bibitem [{\citenamefont {Edwards}\ \emph {et~al.}(2011)\citenamefont
  {Edwards}, \citenamefont {Dudek}, \citenamefont {Richards},\ and\
  \citenamefont {Wallace}}]{Edwards:2011jj}%
  \BibitemOpen
  \bibfield  {author} {\bibinfo {author} {\bibfnamefont {R.~G.}\ \bibnamefont
  {Edwards}}, \bibinfo {author} {\bibfnamefont {J.~J.}\ \bibnamefont {Dudek}},
  \bibinfo {author} {\bibfnamefont {D.~G.}\ \bibnamefont {Richards}}, \ and\
  \bibinfo {author} {\bibfnamefont {S.~J.}\ \bibnamefont {Wallace}},\
  }\bibfield  {title} {\enquote {\bibinfo {title} {{Excited state baryon
  spectroscopy from lattice QCD}},}\ }\href {\doibase
  10.1103/PhysRevD.84.074508} {\bibfield  {journal} {\bibinfo  {journal} {Phys.
  Rev.}\ }\textbf {\bibinfo {volume} {D84}},\ \bibinfo {pages} {074508}
  (\bibinfo {year} {2011})}\BibitemShut {NoStop}%
\bibitem [{\citenamefont {Edwards}\ \emph {et~al.}(2013)\citenamefont
  {Edwards}, \citenamefont {Mathur}, \citenamefont {Richards},\ and\
  \citenamefont {Wallace}}]{Edwards:2012fx}%
  \BibitemOpen
  \bibfield  {author} {\bibinfo {author} {\bibfnamefont {R.~G.}\ \bibnamefont
  {Edwards}}, \bibinfo {author} {\bibfnamefont {N.}~\bibnamefont {Mathur}},
  \bibinfo {author} {\bibfnamefont {D.~G.}\ \bibnamefont {Richards}}, \ and\
  \bibinfo {author} {\bibfnamefont {S.~J.}\ \bibnamefont {Wallace}} (\bibinfo
  {collaboration} {Hadron Spectrum}),\ }\bibfield  {title} {\enquote {\bibinfo
  {title} {{Flavor structure of the excited baryon spectra from lattice
  QCD}},}\ }\href {\doibase 10.1103/PhysRevD.87.054506} {\bibfield  {journal}
  {\bibinfo  {journal} {Phys. Rev.}\ }\textbf {\bibinfo {volume} {D87}},\
  \bibinfo {pages} {054506} (\bibinfo {year} {2013})}\BibitemShut {NoStop}%
\bibitem [{\citenamefont {Capstick}\ and\ \citenamefont
  {Roberts}(2000)}]{Capstick:2000qj}%
  \BibitemOpen
  \bibfield  {author} {\bibinfo {author} {\bibfnamefont {S.}~\bibnamefont
  {Capstick}}\ and\ \bibinfo {author} {\bibfnamefont {W.}~\bibnamefont
  {Roberts}},\ }\bibfield  {title} {\enquote {\bibinfo {title} {{Quark models
  of baryon masses and decays}},}\ }\href@noop {} {\bibfield  {journal}
  {\bibinfo  {journal} {Prog. Part. Nucl. Phys.}\ }\textbf {\bibinfo {volume}
  {45}},\ \bibinfo {pages} {S241--S331} (\bibinfo {year} {2000})}\BibitemShut
  {NoStop}%
\bibitem [{\citenamefont {Capstick}\ and\ \citenamefont
  {Roberts}(1998)}]{Capstick:1998uh}%
  \BibitemOpen
  \bibfield  {author} {\bibinfo {author} {\bibfnamefont {S.}~\bibnamefont
  {Capstick}}\ and\ \bibinfo {author} {\bibfnamefont {W.}~\bibnamefont
  {Roberts}},\ }\bibfield  {title} {\enquote {\bibinfo {title} {{Strange decays
  of nonstrange baryons}},}\ }\href {\doibase 10.1103/PhysRevD.58.074011}
  {\bibfield  {journal} {\bibinfo  {journal} {Phys. Rev.}\ }\textbf {\bibinfo
  {volume} {D58}},\ \bibinfo {pages} {074011} (\bibinfo {year}
  {1998})}\BibitemShut {NoStop}%
\bibitem [{\citenamefont {Capstick}\ and\ \citenamefont
  {Isgur}(1986)}]{Capstick:1986bm}%
  \BibitemOpen
  \bibfield  {author} {\bibinfo {author} {\bibfnamefont {S.}~\bibnamefont
  {Capstick}}\ and\ \bibinfo {author} {\bibfnamefont {N.}~\bibnamefont
  {Isgur}},\ }\bibfield  {title} {\enquote {\bibinfo {title} {{Baryons in a
  Relativized Quark Model with Chromodynamics}},}\ }\bibfield  {booktitle}
  {\emph {\bibinfo {booktitle} {{Proceedings, International Conference on
  Hadron Spectroscopy: College Park, Maryland, April 20-22, 1985}}},\ }\href
  {\doibase 10.1103/PhysRevD.34.2809, 10.1063/1.35361} {\bibfield  {journal}
  {\bibinfo  {journal} {Phys. Rev.}\ }\textbf {\bibinfo {volume} {D34}},\
  \bibinfo {pages} {2809} (\bibinfo {year} {1986})},\ \bibinfo {note} {[AIP
  Conf. Proc. 132, 267 (1985)]}\BibitemShut {NoStop}%
\bibitem [{\citenamefont {Loring}\ \emph {et~al.}(2001)\citenamefont {Loring},
  \citenamefont {M.},\ and\ \citenamefont {Petry}}]{Loring:2001kx}%
  \BibitemOpen
  \bibfield  {author} {\bibinfo {author} {\bibfnamefont {U.}~\bibnamefont
  {Loring}}, \bibinfo {author} {\bibfnamefont {Bernard~C.}\ \bibnamefont {M.}},
  \ and\ \bibinfo {author} {\bibfnamefont {H.~R.}\ \bibnamefont {Petry}},\
  }\bibfield  {title} {\enquote {\bibinfo {title} {{The Light baryon spectrum
  in a relativistic quark model with instanton induced quark forces: The
  Nonstrange baryon spectrum and ground states}},}\ }\href {\doibase
  10.1007/s100500170105} {\bibfield  {journal} {\bibinfo  {journal} {Eur. Phys.
  J.}\ }\textbf {\bibinfo {volume} {A10}},\ \bibinfo {pages} {395--446}
  (\bibinfo {year} {2001})}\BibitemShut {NoStop}%
\bibitem [{\citenamefont {Glozman}\ \emph {et~al.}(1998)\citenamefont
  {Glozman}, \citenamefont {Plessas}, \citenamefont {Varga},\ and\
  \citenamefont {Wagenbrunn}}]{Glozman:1997ag}%
  \BibitemOpen
  \bibfield  {author} {\bibinfo {author} {\bibfnamefont {L.~{\relax Ya}.}\
  \bibnamefont {Glozman}}, \bibinfo {author} {\bibfnamefont {W.}~\bibnamefont
  {Plessas}}, \bibinfo {author} {\bibfnamefont {K.}~\bibnamefont {Varga}}, \
  and\ \bibinfo {author} {\bibfnamefont {R.~F.}\ \bibnamefont {Wagenbrunn}},\
  }\bibfield  {title} {\enquote {\bibinfo {title} {{Unified description of
  light and strange baryon spectra}},}\ }\href {\doibase
  10.1103/PhysRevD.58.094030} {\bibfield  {journal} {\bibinfo  {journal} {Phys.
  Rev.}\ }\textbf {\bibinfo {volume} {D58}},\ \bibinfo {pages} {094030}
  (\bibinfo {year} {1998})}\BibitemShut {NoStop}%
\bibitem [{\citenamefont {Giannini}\ \emph {et~al.}(2001)\citenamefont
  {Giannini}, \citenamefont {Santopinto},\ and\ \citenamefont
  {Vassallo}}]{Giannini:2001kb}%
  \BibitemOpen
  \bibfield  {author} {\bibinfo {author} {\bibfnamefont {M.~M.}\ \bibnamefont
  {Giannini}}, \bibinfo {author} {\bibfnamefont {E.}~\bibnamefont
  {Santopinto}}, \ and\ \bibinfo {author} {\bibfnamefont {A.}~\bibnamefont
  {Vassallo}},\ }\bibfield  {title} {\enquote {\bibinfo {title} {{Hypercentral
  constituent quark model and isospin dependence}},}\ }\href {\doibase
  10.1007/s10050-001-8668-y} {\bibfield  {journal} {\bibinfo  {journal} {Eur.
  Phys. J.}\ }\textbf {\bibinfo {volume} {A12}},\ \bibinfo {pages} {447--452}
  (\bibinfo {year} {2001})}\BibitemShut {NoStop}%
\bibitem [{\citenamefont {Patrignani}\ \emph {et~al.}(2016)\citenamefont
  {Patrignani} \emph {et~al.}}]{Patrignani:2016xqp}%
  \BibitemOpen
  \bibfield  {author} {\bibinfo {author} {\bibfnamefont {C.}~\bibnamefont
  {Patrignani}} \emph {et~al.} (\bibinfo {collaboration} {Particle Data
  Group}),\ }\bibfield  {title} {\enquote {\bibinfo {title} {{Review of
  Particle Physics}},}\ }\href {\doibase 10.1088/1674-1137/40/10/100001}
  {\bibfield  {journal} {\bibinfo  {journal} {Chin. Phys.}\ }\textbf {\bibinfo
  {volume} {C40}},\ \bibinfo {pages} {100001} (\bibinfo {year}
  {2016})}\BibitemShut {NoStop}%
\bibitem [{\citenamefont {Klempt}\ and\ \citenamefont
  {Richard}(2010)}]{Klempt:2009pi}%
  \BibitemOpen
  \bibfield  {author} {\bibinfo {author} {\bibfnamefont {E.}~\bibnamefont
  {Klempt}}\ and\ \bibinfo {author} {\bibfnamefont {J-M.}\ \bibnamefont
  {Richard}},\ }\bibfield  {title} {\enquote {\bibinfo {title} {{Baryon
  spectroscopy}},}\ }\href {\doibase 10.1103/RevModPhys.82.1095} {\bibfield
  {journal} {\bibinfo  {journal} {Rev. Mod. Phys.}\ }\textbf {\bibinfo {volume}
  {82}},\ \bibinfo {pages} {1095--1153} (\bibinfo {year} {2010})}\BibitemShut
  {NoStop}%
\bibitem [{\citenamefont {Koniuk}\ and\ \citenamefont
  {Isgur}(1980)}]{Koniuk:1979vw}%
  \BibitemOpen
  \bibfield  {author} {\bibinfo {author} {\bibfnamefont {R.}~\bibnamefont
  {Koniuk}}\ and\ \bibinfo {author} {\bibfnamefont {N.}~\bibnamefont {Isgur}},\
  }\bibfield  {title} {\enquote {\bibinfo {title} {{Where Have All the
  Resonances Gone? An Analysis of Baryon Couplings in a Quark Model With
  Chromodynamics}},}\ }\bibfield  {booktitle} {\emph {\bibinfo {booktitle}
  {{Baryon 1980:217}}},\ }\href {\doibase 10.1103/PhysRevLett.44.845}
  {\bibfield  {journal} {\bibinfo  {journal} {Phys. Rev. Lett.}\ }\textbf
  {\bibinfo {volume} {44}},\ \bibinfo {pages} {845} (\bibinfo {year}
  {1980})}\BibitemShut {NoStop}%
\bibitem [{\citenamefont {Anselmino}\ \emph {et~al.}(1993)\citenamefont
  {Anselmino}, \citenamefont {Predazzi}, \citenamefont {Ekelin}, \citenamefont
  {Fredriksson},\ and\ \citenamefont {Lichtenberg}}]{Anselmino:1992vg}%
  \BibitemOpen
  \bibfield  {author} {\bibinfo {author} {\bibfnamefont {M.}~\bibnamefont
  {Anselmino}}, \bibinfo {author} {\bibfnamefont {E.}~\bibnamefont {Predazzi}},
  \bibinfo {author} {\bibfnamefont {S.}~\bibnamefont {Ekelin}}, \bibinfo
  {author} {\bibfnamefont {S.}~\bibnamefont {Fredriksson}}, \ and\ \bibinfo
  {author} {\bibfnamefont {D.~B.}\ \bibnamefont {Lichtenberg}},\ }\bibfield
  {title} {\enquote {\bibinfo {title} {{Diquarks}},}\ }\href {\doibase
  10.1103/RevModPhys.65.1199} {\bibfield  {journal} {\bibinfo  {journal} {Rev.
  Mod. Phys.}\ }\textbf {\bibinfo {volume} {65}},\ \bibinfo {pages}
  {1199--1234} (\bibinfo {year} {1993})}\BibitemShut {NoStop}%
\bibitem [{\citenamefont {Brodsky}(2007)}]{Brodsky:2006uq}%
  \BibitemOpen
  \bibfield  {author} {\bibinfo {author} {\bibfnamefont {S.~J.}\ \bibnamefont
  {Brodsky}},\ }\bibfield  {title} {\enquote {\bibinfo {title} {{Hadron
  Spectroscopy and Structure from AdS/CFT}},}\ }\href {\doibase
  10.1140/epja/i2006-10221-7} {\bibfield  {journal} {\bibinfo  {journal} {Eur.
  Phys. J.}\ }\textbf {\bibinfo {volume} {A31}},\ \bibinfo {pages} {638--644}
  (\bibinfo {year} {2007})}\BibitemShut {NoStop}%
\bibitem [{\citenamefont {Kolomeitsev}\ and\ \citenamefont
  {Lutz}(2004)}]{Kolomeitsev:2003kt}%
  \BibitemOpen
  \bibfield  {author} {\bibinfo {author} {\bibfnamefont {E.~E.}\ \bibnamefont
  {Kolomeitsev}}\ and\ \bibinfo {author} {\bibfnamefont {M.~F.~M.}\
  \bibnamefont {Lutz}},\ }\bibfield  {title} {\enquote {\bibinfo {title} {{On
  baryon resonances and chiral symmetry}},}\ }\href {\doibase
  10.1016/j.physletb.2004.01.066} {\bibfield  {journal} {\bibinfo  {journal}
  {Phys. Lett.}\ }\textbf {\bibinfo {volume} {B585}},\ \bibinfo {pages}
  {243--252} (\bibinfo {year} {2004})}\BibitemShut {NoStop}%
\bibitem [{\citenamefont {Chew}\ \emph {et~al.}(1957)\citenamefont {Chew},
  \citenamefont {Goldberger}, \citenamefont {Low},\ and\ \citenamefont
  {Nambu}}]{Chew:1957tf}%
  \BibitemOpen
  \bibfield  {author} {\bibinfo {author} {\bibfnamefont {G.~F.}\ \bibnamefont
  {Chew}}, \bibinfo {author} {\bibfnamefont {M.~L.}\ \bibnamefont
  {Goldberger}}, \bibinfo {author} {\bibfnamefont {F.~E.}\ \bibnamefont {Low}},
  \ and\ \bibinfo {author} {\bibfnamefont {Yoichiro}\ \bibnamefont {Nambu}},\
  }\bibfield  {title} {\enquote {\bibinfo {title} {{Relativistic dispersion
  relation approach to photomeson production}},}\ }\href {\doibase
  10.1103/PhysRev.106.1345} {\bibfield  {journal} {\bibinfo  {journal} {Phys.
  Rev.}\ }\textbf {\bibinfo {volume} {106}},\ \bibinfo {pages} {1345--1355}
  (\bibinfo {year} {1957})}\BibitemShut {NoStop}%
\bibitem [{\citenamefont {Barker}\ \emph {et~al.}(1975)\citenamefont {Barker},
  \citenamefont {Donnachie},\ and\ \citenamefont {Storrow}}]{Barker:1975bp}%
  \BibitemOpen
  \bibfield  {author} {\bibinfo {author} {\bibfnamefont {I.~S.}\ \bibnamefont
  {Barker}}, \bibinfo {author} {\bibfnamefont {A.}~\bibnamefont {Donnachie}}, \
  and\ \bibinfo {author} {\bibfnamefont {J.~K.}\ \bibnamefont {Storrow}},\
  }\bibfield  {title} {\enquote {\bibinfo {title} {{Complete Experiments in
  Pseudoscalar Photoproduction}},}\ }\href {\doibase
  10.1016/0550-3213(75)90049-8} {\bibfield  {journal} {\bibinfo  {journal}
  {Nucl. Phys.}\ }\textbf {\bibinfo {volume} {B95}},\ \bibinfo {pages}
  {347--356} (\bibinfo {year} {1975})}\BibitemShut {NoStop}%
\bibitem [{\citenamefont {Fasano}\ \emph {et~al.}(1992)\citenamefont {Fasano},
  \citenamefont {Tabakin},\ and\ \citenamefont {Saghai}}]{Fasano:1992es}%
  \BibitemOpen
  \bibfield  {author} {\bibinfo {author} {\bibfnamefont {C.~G.}\ \bibnamefont
  {Fasano}}, \bibinfo {author} {\bibfnamefont {F.}~\bibnamefont {Tabakin}}, \
  and\ \bibinfo {author} {\bibfnamefont {B.}~\bibnamefont {Saghai}},\
  }\bibfield  {title} {\enquote {\bibinfo {title} {{Spin observables at
  threshold for meson photoproduction}},}\ }\href {\doibase
  10.1103/PhysRevC.46.2430} {\bibfield  {journal} {\bibinfo  {journal} {Phys.
  Rev.}\ }\textbf {\bibinfo {volume} {C46}},\ \bibinfo {pages} {2430--2455}
  (\bibinfo {year} {1992})}\BibitemShut {NoStop}%
\bibitem [{\citenamefont {Chiang}\ and\ \citenamefont
  {Tabakin}(1997)}]{Chiang:1996em}%
  \BibitemOpen
  \bibfield  {author} {\bibinfo {author} {\bibfnamefont {W-T.}\ \bibnamefont
  {Chiang}}\ and\ \bibinfo {author} {\bibfnamefont {F.}~\bibnamefont
  {Tabakin}},\ }\bibfield  {title} {\enquote {\bibinfo {title} {{Completeness
  rules for spin observables in pseudoscalar meson photoproduction}},}\ }\href
  {\doibase 10.1103/PhysRevC.55.2054} {\bibfield  {journal} {\bibinfo
  {journal} {Phys. Rev.}\ }\textbf {\bibinfo {volume} {C55}},\ \bibinfo {pages}
  {2054--2066} (\bibinfo {year} {1997})}\BibitemShut {NoStop}%
\bibitem [{\citenamefont {Keaton}\ and\ \citenamefont
  {Workman}(1996)}]{Keaton:1996pe}%
  \BibitemOpen
  \bibfield  {author} {\bibinfo {author} {\bibfnamefont {G.}~\bibnamefont
  {Keaton}}\ and\ \bibinfo {author} {\bibfnamefont {R.}~\bibnamefont
  {Workman}},\ }\bibfield  {title} {\enquote {\bibinfo {title} {{Ambiguities in
  the partial wave analysis of pseudoscalar meson photoproduction}},}\ }\href
  {\doibase 10.1103/PhysRevC.54.1437} {\bibfield  {journal} {\bibinfo
  {journal} {Phys. Rev.}\ }\textbf {\bibinfo {volume} {C54}},\ \bibinfo {pages}
  {1437--1440} (\bibinfo {year} {1996})}\BibitemShut {NoStop}%
\bibitem [{\citenamefont {Sandorfi}\ \emph {et~al.}(2011)\citenamefont
  {Sandorfi}, \citenamefont {Hoblit}, \citenamefont {Kamano},\ and\
  \citenamefont {Lee}}]{Sandorfi:2010uv}%
  \BibitemOpen
  \bibfield  {author} {\bibinfo {author} {\bibfnamefont {A.~M.}\ \bibnamefont
  {Sandorfi}}, \bibinfo {author} {\bibfnamefont {S.}~\bibnamefont {Hoblit}},
  \bibinfo {author} {\bibfnamefont {H.}~\bibnamefont {Kamano}}, \ and\ \bibinfo
  {author} {\bibfnamefont {T.~S.~H.}\ \bibnamefont {Lee}},\ }\bibfield  {title}
  {\enquote {\bibinfo {title} {{Determining pseudoscalar meson photo-production
  amplitudes from complete experiments}},}\ }\href {\doibase
  10.1088/0954-3899/38/5/053001} {\bibfield  {journal} {\bibinfo  {journal} {J.
  Phys.}\ }\textbf {\bibinfo {volume} {G38}},\ \bibinfo {pages} {053001}
  (\bibinfo {year} {2011})}\BibitemShut {NoStop}%
\bibitem [{\citenamefont {Nys}\ \emph {et~al.}(2016)\citenamefont {Nys},
  \citenamefont {Ryckebusch}, \citenamefont {Ireland},\ and\ \citenamefont
  {Glazier}}]{Nys:2016uel}%
  \BibitemOpen
  \bibfield  {author} {\bibinfo {author} {\bibfnamefont {J.}~\bibnamefont
  {Nys}}, \bibinfo {author} {\bibfnamefont {J.}~\bibnamefont {Ryckebusch}},
  \bibinfo {author} {\bibfnamefont {D.~G.}\ \bibnamefont {Ireland}}, \ and\
  \bibinfo {author} {\bibfnamefont {D.~I.}\ \bibnamefont {Glazier}},\
  }\bibfield  {title} {\enquote {\bibinfo {title} {{Model discrimination in
  pseudoscalar-meson photoproduction}},}\ }\href {\doibase
  10.1016/j.physletb.2016.05.069} {\bibfield  {journal} {\bibinfo  {journal}
  {Phys. Lett.}\ }\textbf {\bibinfo {volume} {B759}},\ \bibinfo {pages}
  {260--265} (\bibinfo {year} {2016})}\BibitemShut {NoStop}%
\bibitem [{\citenamefont {Sandorfi}\ and\ \citenamefont
  {Hoblit}(2013)}]{Sandorfi:2013cya}%
  \BibitemOpen
  \bibfield  {author} {\bibinfo {author} {\bibfnamefont {A.~M.}\ \bibnamefont
  {Sandorfi}}\ and\ \bibinfo {author} {\bibfnamefont {S.}~\bibnamefont
  {Hoblit}},\ }\bibfield  {title} {\enquote {\bibinfo {title} {{Hyperon
  photoproduction from polarized H and D towards a $complete$ N$^{*}$
  experiment}},}\ }\bibfield  {booktitle} {\emph {\bibinfo {booktitle}
  {{Proceedings, 11th International Conference on Hypernuclear and Strange
  Particle Physics (HYP 2012): Barcelona, Spain, October 1-5, 2012}}},\ }\href
  {\doibase 10.1016/j.nuclphysa.2013.01.009} {\bibfield  {journal} {\bibinfo
  {journal} {Nucl. Phys.}\ }\textbf {\bibinfo {volume} {A914}},\ \bibinfo
  {pages} {538--542} (\bibinfo {year} {2013})}\BibitemShut {NoStop}%
\bibitem [{\citenamefont {Anisovich}\ \emph
  {et~al.}(2017{\natexlab{a}})\citenamefont {Anisovich}, \citenamefont
  {Burkert}, \citenamefont {Compton}, \citenamefont {Hicks}, \citenamefont
  {Klein}, \citenamefont {Klempt}, \citenamefont {Nikonov}, \citenamefont
  {Sandorfi}, \citenamefont {Sarantsev},\ and\ \citenamefont
  {Thoma}}]{Anisovich:2017afs}%
  \BibitemOpen
  \bibfield  {author} {\bibinfo {author} {\bibfnamefont {A.~V.}\ \bibnamefont
  {Anisovich}}, \bibinfo {author} {\bibfnamefont {V.}~\bibnamefont {Burkert}},
  \bibinfo {author} {\bibfnamefont {N.}~\bibnamefont {Compton}}, \bibinfo
  {author} {\bibfnamefont {K.}~\bibnamefont {Hicks}}, \bibinfo {author}
  {\bibfnamefont {F.~J.}\ \bibnamefont {Klein}}, \bibinfo {author}
  {\bibfnamefont {E.}~\bibnamefont {Klempt}}, \bibinfo {author} {\bibfnamefont
  {V.~A.}\ \bibnamefont {Nikonov}}, \bibinfo {author} {\bibfnamefont {A.~M.}\
  \bibnamefont {Sandorfi}}, \bibinfo {author} {\bibfnamefont {A.~V.}\
  \bibnamefont {Sarantsev}}, \ and\ \bibinfo {author} {\bibfnamefont
  {U.}~\bibnamefont {Thoma}},\ }\bibfield  {title} {\enquote {\bibinfo {title}
  {{Neutron helicity amplitudes}},}\ }\href {\doibase
  10.1103/PhysRevC.96.055202} {\bibfield  {journal} {\bibinfo  {journal} {Phys.
  Rev.}\ }\textbf {\bibinfo {volume} {C 96}},\ \bibinfo {pages} {055202}
  (\bibinfo {year} {2017}{\natexlab{a}})}\BibitemShut {NoStop}%
\bibitem [{\citenamefont {Compton}\ \emph {et~al.}(2017)\citenamefont {Compton}
  \emph {et~al.}}]{Compton_PhysRevC.96.065201}%
  \BibitemOpen
  \bibfield  {author} {\bibinfo {author} {\bibfnamefont {N.}~\bibnamefont
  {Compton}} \emph {et~al.} (\bibinfo {collaboration} {CLAS Collaboration}),\
  }\bibfield  {title} {\enquote {\bibinfo {title} {Measurement of the
  differential and total cross sections of the
  $\ensuremath{\gamma}d\ensuremath{\rightarrow}{K}^{0}\mathrm{\ensuremath{\Lambda}}(p)$
  reaction within the resonance region},}\ }\href {\doibase
  10.1103/PhysRevC.96.065201} {\bibfield  {journal} {\bibinfo  {journal} {Phys.
  Rev. C}\ }\textbf {\bibinfo {volume} {96}},\ \bibinfo {pages} {065201}
  (\bibinfo {year} {2017})}\BibitemShut {NoStop}%
\bibitem [{\citenamefont {Anafelos~Pereira}\ \emph {et~al.}(2010)\citenamefont
  {Anafelos~Pereira} \emph {et~al.}}]{AnefalosPereira:2009zw}%
  \BibitemOpen
  \bibfield  {author} {\bibinfo {author} {\bibfnamefont {S.}~\bibnamefont
  {Anafelos~Pereira}} \emph {et~al.} (\bibinfo {collaboration} {CLAS
  Collaboration}),\ }\bibfield  {title} {\enquote {\bibinfo {title}
  {{Differential cross section of $\gamma n \to K^+ \Sigma^-$ on bound neutrons
  with incident photons from 1.1 to 3.6 GeV}},}\ }\href {\doibase
  10.1016/j.physletb.2010.04.028} {\bibfield  {journal} {\bibinfo  {journal}
  {Phys. Lett.}\ }\textbf {\bibinfo {volume} {B688}},\ \bibinfo {pages}
  {289--293} (\bibinfo {year} {2010})}\BibitemShut {NoStop}%
\bibitem [{\citenamefont {Paterson}\ \emph {et~al.}(2016)\citenamefont
  {Paterson} \emph {et~al.}}]{Paterson:2016vmc}%
  \BibitemOpen
  \bibfield  {author} {\bibinfo {author} {\bibfnamefont {C.~A.}\ \bibnamefont
  {Paterson}} \emph {et~al.} (\bibinfo {collaboration} {CLAS Collaboration}),\
  }\bibfield  {title} {\enquote {\bibinfo {title} {{Photoproduction of
  $\Lambda$ and $\Sigma^0$ hyperons using linearly polarized photons}},}\
  }\href {\doibase 10.1103/PhysRevC.93.065201} {\bibfield  {journal} {\bibinfo
  {journal} {Phys. Rev.}\ }\textbf {\bibinfo {volume} {C93}},\ \bibinfo {pages}
  {065201} (\bibinfo {year} {2016})}\BibitemShut {NoStop}%
\bibitem [{\citenamefont {Bradford}\ \emph {et~al.}(2007)\citenamefont
  {Bradford} \emph {et~al.}}]{Bradford_CxCz}%
  \BibitemOpen
  \bibfield  {author} {\bibinfo {author} {\bibfnamefont {R.}~\bibnamefont
  {Bradford}} \emph {et~al.} (\bibinfo {collaboration} {CLAS Collaboration}),\
  }\bibfield  {title} {\enquote {\bibinfo {title} {{First measurement of
  beam-recoil observables $C_{x}$ and $C_{z}$ in hyperon photoproduction}},}\
  }\href {\doibase 10.1103/PhysRevC.75.035205} {\bibfield  {journal} {\bibinfo
  {journal} {Phys. Rev.}\ }\textbf {\bibinfo {volume} {C75}},\ \bibinfo {pages}
  {035205} (\bibinfo {year} {2007})}\BibitemShut {NoStop}%
\bibitem [{\citenamefont {Bradford}\ \emph {et~al.}(2006)\citenamefont
  {Bradford} \emph {et~al.}}]{Bradford_xsec}%
  \BibitemOpen
  \bibfield  {author} {\bibinfo {author} {\bibfnamefont {R.}~\bibnamefont
  {Bradford}} \emph {et~al.} (\bibinfo {collaboration} {CLAS Collaboration}),\
  }\bibfield  {title} {\enquote {\bibinfo {title} {{Differential cross sections
  for $\gamma + p \to K^+ + Y$ for $\Lambda$ and $\Sigma^0$ hyperons}},}\
  }\href {\doibase 10.1103/PhysRevC.73.035202} {\bibfield  {journal} {\bibinfo
  {journal} {Phys. Rev.}\ }\textbf {\bibinfo {volume} {C73}},\ \bibinfo {pages}
  {035202} (\bibinfo {year} {2006})}\BibitemShut {NoStop}%
\bibitem [{\citenamefont {McNabb}\ \emph {et~al.}(2004)\citenamefont {McNabb}
  \emph {et~al.}}]{McNabb}%
  \BibitemOpen
  \bibfield  {author} {\bibinfo {author} {\bibfnamefont {J.~W.~C.}\
  \bibnamefont {McNabb}} \emph {et~al.} (\bibinfo {collaboration} {CLAS
  Collaboration}),\ }\bibfield  {title} {\enquote {\bibinfo {title} {{Hyperon
  photoproduction in the nucleon resonance region}},}\ }\href {\doibase
  10.1103/PhysRevC.69.042201} {\bibfield  {journal} {\bibinfo  {journal} {Phys.
  Rev.}\ }\textbf {\bibinfo {volume} {C69}},\ \bibinfo {pages} {042201}
  (\bibinfo {year} {2004})}\BibitemShut {NoStop}%
\bibitem [{\citenamefont {Anisovich}\ \emph {et~al.}(2007)\citenamefont
  {Anisovich}, \citenamefont {Kleber}, \citenamefont {Klempt}, \citenamefont
  {Nikonov}, \citenamefont {Sarantsev},\ and\ \citenamefont
  {Thoma}}]{Anisovich:2007bq}%
  \BibitemOpen
  \bibfield  {author} {\bibinfo {author} {\bibfnamefont {A.~V.}\ \bibnamefont
  {Anisovich}}, \bibinfo {author} {\bibfnamefont {V.}~\bibnamefont {Kleber}},
  \bibinfo {author} {\bibfnamefont {E.}~\bibnamefont {Klempt}}, \bibinfo
  {author} {\bibfnamefont {V.~A.}\ \bibnamefont {Nikonov}}, \bibinfo {author}
  {\bibfnamefont {A.~V.}\ \bibnamefont {Sarantsev}}, \ and\ \bibinfo {author}
  {\bibfnamefont {U.}~\bibnamefont {Thoma}},\ }\bibfield  {title} {\enquote
  {\bibinfo {title} {{Baryon resonances and polarization transfer in hyperon
  photoproduction}},}\ }\href {\doibase 10.1140/epja/i2007-10503-6} {\bibfield
  {journal} {\bibinfo  {journal} {Eur. Phys. J.}\ }\textbf {\bibinfo {volume}
  {A34}},\ \bibinfo {pages} {243--254} (\bibinfo {year} {2007})}\BibitemShut
  {NoStop}%
\bibitem [{\citenamefont {McCracken}\ \emph {et~al.}(2010)\citenamefont
  {McCracken} \emph {et~al.}}]{McCracken}%
  \BibitemOpen
  \bibfield  {author} {\bibinfo {author} {\bibfnamefont {M.~E.}\ \bibnamefont
  {McCracken}} \emph {et~al.} (\bibinfo {collaboration} {CLAS Collaboration}),\
  }\bibfield  {title} {\enquote {\bibinfo {title} {{Differential cross section
  and recoil polarization measurements for the $\gamma p \to K^+ \Lambda$
  reaction using CLAS at Jefferson Lab}},}\ }\href {\doibase
  10.1103/PhysRevC.81.025201} {\bibfield  {journal} {\bibinfo  {journal} {Phys.
  Rev.}\ }\textbf {\bibinfo {volume} {C81}},\ \bibinfo {pages} {025201}
  (\bibinfo {year} {2010})}\BibitemShut {NoStop}%
\bibitem [{\citenamefont {Dey}\ \emph {et~al.}(2010)\citenamefont {Dey} \emph
  {et~al.}}]{Dey}%
  \BibitemOpen
  \bibfield  {author} {\bibinfo {author} {\bibfnamefont {B.}~\bibnamefont
  {Dey}} \emph {et~al.} (\bibinfo {collaboration} {CLAS Collaboration}),\
  }\bibfield  {title} {\enquote {\bibinfo {title} {{Differential cross sections
  and recoil polarizations for the reaction $\gamma p \to K^+ \Sigma^0$}},}\
  }\href@noop {} {\bibfield  {journal} {\bibinfo  {journal} {Phys. Rev.}\
  }\textbf {\bibinfo {volume} {C82}},\ \bibinfo {pages} {025202} (\bibinfo
  {year} {2010})}\BibitemShut {NoStop}%
\bibitem [{\citenamefont {Anisovich}\ \emph
  {et~al.}(2017{\natexlab{b}})\citenamefont {Anisovich} \emph
  {et~al.}}]{Anisovich:2017ygb}%
  \BibitemOpen
  \bibfield  {author} {\bibinfo {author} {\bibfnamefont {A.~V.}\ \bibnamefont
  {Anisovich}} \emph {et~al.},\ }\bibfield  {title} {\enquote {\bibinfo {title}
  {{$N^*$ resonances from $K\Lambda$ amplitudes in sliced bins in energy}},}\
  }\href {\doibase 10.1140/epja/i2017-12443-x} {\bibfield  {journal} {\bibinfo
  {journal} {Eur. Phys. J.}\ }\textbf {\bibinfo {volume} {A53}},\ \bibinfo
  {pages} {242} (\bibinfo {year} {2017}{\natexlab{b}})}\BibitemShut {NoStop}%
\bibitem [{\citenamefont {Mart}\ and\ \citenamefont
  {Bennhold}(1999)}]{Mart:1999ed}%
  \BibitemOpen
  \bibfield  {author} {\bibinfo {author} {\bibfnamefont {T.}~\bibnamefont
  {Mart}}\ and\ \bibinfo {author} {\bibfnamefont {C.}~\bibnamefont
  {Bennhold}},\ }\bibfield  {title} {\enquote {\bibinfo {title} {{Evidence for
  a missing nucleon resonance in kaon photoproduction}},}\ }\href {\doibase
  10.1103/PhysRevC.61.012201} {\bibfield  {journal} {\bibinfo  {journal} {Phys.
  Rev.}\ }\textbf {\bibinfo {volume} {C61}},\ \bibinfo {pages} {012201}
  (\bibinfo {year} {1999})}\BibitemShut {NoStop}%
\bibitem [{\citenamefont {Lee}\ \emph {et~al.}(2001)\citenamefont {Lee},
  \citenamefont {Mart}, \citenamefont {Bennhold},\ and\ \citenamefont
  {Wright}}]{Lee:1999kd}%
  \BibitemOpen
  \bibfield  {author} {\bibinfo {author} {\bibfnamefont {F.~X.}\ \bibnamefont
  {Lee}}, \bibinfo {author} {\bibfnamefont {T.}~\bibnamefont {Mart}}, \bibinfo
  {author} {\bibfnamefont {C.}~\bibnamefont {Bennhold}}, \ and\ \bibinfo
  {author} {\bibfnamefont {L.~E.}\ \bibnamefont {Wright}},\ }\bibfield  {title}
  {\enquote {\bibinfo {title} {{Quasifree kaon photoproduction on nuclei}},}\
  }\href {\doibase 10.1016/S0375-9474(01)01098-3} {\bibfield  {journal}
  {\bibinfo  {journal} {Nucl. Phys.}\ }\textbf {\bibinfo {volume} {A695}},\
  \bibinfo {pages} {237--272} (\bibinfo {year} {2001})}\BibitemShut {NoStop}%
\bibitem [{\citenamefont {Arndt}\ \emph {et~al.}(2015)\citenamefont {Arndt},
  \citenamefont {Briscoe}, \citenamefont {Strakovsky},\ and\ \citenamefont
  {Workman}}]{SAID}%
  \BibitemOpen
  \bibfield  {author} {\bibinfo {author} {\bibfnamefont {R.}~\bibnamefont
  {Arndt}}, \bibinfo {author} {\bibfnamefont {W.}~\bibnamefont {Briscoe}},
  \bibinfo {author} {\bibfnamefont {I.}~\bibnamefont {Strakovsky}}, \ and\
  \bibinfo {author} {\bibfnamefont {R.}~\bibnamefont {Workman}} (\bibinfo
  {collaboration} {{George Washington Data Analysis Center}}),\ }\href@noop {}
  {\enquote {\bibinfo {title} {{SAID}},}\ } (\bibinfo {year} {2015}),\ \bibinfo
  {note} {{SAID web site \url{http://gwdac.phys.gwu.edu}}}\BibitemShut
  {NoStop}%
\bibitem [{\citenamefont {Adelseck}\ \emph {et~al.}(1985)\citenamefont
  {Adelseck}, \citenamefont {Bennhold},\ and\ \citenamefont
  {Wright}}]{Adelseck:1986fb}%
  \BibitemOpen
  \bibfield  {author} {\bibinfo {author} {\bibfnamefont {R.~A.}\ \bibnamefont
  {Adelseck}}, \bibinfo {author} {\bibfnamefont {C.}~\bibnamefont {Bennhold}},
  \ and\ \bibinfo {author} {\bibfnamefont {L.~E.}\ \bibnamefont {Wright}},\
  }\bibfield  {title} {\enquote {\bibinfo {title} {{Kaon Photoproduction
  Operator for Use in Nuclear Physics}},}\ }\href {\doibase
  10.1103/PhysRevC.32.1681} {\bibfield  {journal} {\bibinfo  {journal} {Phys.
  Rev.}\ }\textbf {\bibinfo {volume} {C32}},\ \bibinfo {pages} {1681--1692}
  (\bibinfo {year} {1985})}\BibitemShut {NoStop}%
\bibitem [{\citenamefont {Strakovsky}(2017)}]{strakovsky}%
  \BibitemOpen
  \bibfield  {author} {\bibinfo {author} {\bibfnamefont {I.}~\bibnamefont
  {Strakovsky}} (\bibinfo {collaboration} {{George Washington Data Analysis
  Center}}),\ }\href@noop {} {\enquote {\bibinfo {title} {{SAID}},}\ }
  (\bibinfo {year} {2017}),\ \bibinfo {note} {{Private
  Communication}}\BibitemShut {NoStop}%
\bibitem [{\citenamefont {Anisovich}\ \emph {et~al.}(2012)\citenamefont
  {Anisovich}, \citenamefont {Beck}, \citenamefont {Klempt}, \citenamefont
  {Nikonov}, \citenamefont {Sarantsev},\ and\ \citenamefont
  {Thoma}}]{Anisovich:2012ct}%
  \BibitemOpen
  \bibfield  {author} {\bibinfo {author} {\bibfnamefont {A.~V.}\ \bibnamefont
  {Anisovich}}, \bibinfo {author} {\bibfnamefont {R.}~\bibnamefont {Beck}},
  \bibinfo {author} {\bibfnamefont {E.}~\bibnamefont {Klempt}}, \bibinfo
  {author} {\bibfnamefont {V.~A.}\ \bibnamefont {Nikonov}}, \bibinfo {author}
  {\bibfnamefont {A.~V.}\ \bibnamefont {Sarantsev}}, \ and\ \bibinfo {author}
  {\bibfnamefont {U.}~\bibnamefont {Thoma}},\ }\bibfield  {title} {\enquote
  {\bibinfo {title} {{Pion- and photo-induced transition amplitudes to $\Lambda
  K$, $\Sigma K$, and $N\eta$}},}\ }\href {\doibase 10.1140/epja/i2012-12088-3}
  {\bibfield  {journal} {\bibinfo  {journal} {Eur. Phys. J.}\ }\textbf
  {\bibinfo {volume} {A48}},\ \bibinfo {pages} {88} (\bibinfo {year}
  {2012})}\BibitemShut {NoStop}%
\bibitem [{\citenamefont {Mecking}\ \emph {et~al.}(2003)\citenamefont {Mecking}
  \emph {et~al.}}]{CLAS-NIM}%
  \BibitemOpen
  \bibfield  {author} {\bibinfo {author} {\bibfnamefont {B.~A.}\ \bibnamefont
  {Mecking}} \emph {et~al.},\ }\bibfield  {title} {\enquote {\bibinfo {title}
  {{The CEBAF Large Acceptance Spectrometer (CLAS)}},}\ }\href {\doibase
  10.1016/S0168-9002(03)01001-5} {\bibfield  {journal} {\bibinfo  {journal}
  {Nucl. Instrum. Meth.}\ }\textbf {\bibinfo {volume} {A503}},\ \bibinfo
  {pages} {513--553} (\bibinfo {year} {2003})}\BibitemShut {NoStop}%
\bibitem [{\citenamefont {Moriya}\ \emph {et~al.}(2014)\citenamefont {Moriya}
  \emph {et~al.}}]{Moriya:2014kpv}%
  \BibitemOpen
  \bibfield  {author} {\bibinfo {author} {\bibfnamefont {K.}~\bibnamefont
  {Moriya}} \emph {et~al.} (\bibinfo {collaboration} {CLAS Collaboration}),\
  }\bibfield  {title} {\enquote {\bibinfo {title} {{Spin and parity measurement
  of the $\Lambda(1405)$ baryon}},}\ }\href {\doibase
  10.1103/PhysRevLett.112.082004} {\bibfield  {journal} {\bibinfo  {journal}
  {Phys. Rev. Lett.}\ }\textbf {\bibinfo {volume} {112}},\ \bibinfo {pages}
  {082004} (\bibinfo {year} {2014})}\BibitemShut {NoStop}%
\bibitem [{\citenamefont {Moriya}\ \emph
  {et~al.}(2013{\natexlab{a}})\citenamefont {Moriya} \emph
  {et~al.}}]{Moriya:2013hwg}%
  \BibitemOpen
  \bibfield  {author} {\bibinfo {author} {\bibfnamefont {K.}~\bibnamefont
  {Moriya}} \emph {et~al.} (\bibinfo {collaboration} {CLAS Collaboration}),\
  }\bibfield  {title} {\enquote {\bibinfo {title} {{Differential
  Photoproduction Cross Sections of the $\Sigma^0(1385)$, $\Lambda(1405)$, and
  $\Lambda(1520)$}},}\ }\href {\doibase 10.1103/PhysRevC.88.049902,
  10.1103/PhysRevC.88.045201} {\bibfield  {journal} {\bibinfo  {journal} {Phys.
  Rev.}\ }\textbf {\bibinfo {volume} {C 88}},\ \bibinfo {pages} {045201}
  (\bibinfo {year} {2013}{\natexlab{a}})},\ \bibinfo {note} {[Addendum: Phys.
  Rev.C88,no.4,049902(2013)]}\BibitemShut {NoStop}%
\bibitem [{\citenamefont {Moriya}\ \emph
  {et~al.}(2013{\natexlab{b}})\citenamefont {Moriya} \emph
  {et~al.}}]{Moriya:2013eb}%
  \BibitemOpen
  \bibfield  {author} {\bibinfo {author} {\bibfnamefont {K.}~\bibnamefont
  {Moriya}} \emph {et~al.} (\bibinfo {collaboration} {CLAS Collaboration}),\
  }\bibfield  {title} {\enquote {\bibinfo {title} {{Measurement of the
  $\Sigma\pi$ photoproduction line shapes near the $\Lambda(1405)$}},}\ }\href
  {\doibase 10.1103/PhysRevC.87.035206} {\bibfield  {journal} {\bibinfo
  {journal} {Phys. Rev.}\ }\textbf {\bibinfo {volume} {C87}},\ \bibinfo {pages}
  {035206} (\bibinfo {year} {2013}{\natexlab{b}})}\BibitemShut {NoStop}%
\bibitem [{\citenamefont {Ho}\ \emph {et~al.}(2017)\citenamefont {Ho} \emph
  {et~al.}}]{Ho:2017kca}%
  \BibitemOpen
  \bibfield  {author} {\bibinfo {author} {\bibfnamefont {D.}~\bibnamefont {Ho}}
  \emph {et~al.} (\bibinfo {collaboration} {CLAS Collaboration}),\ }\bibfield
  {title} {\enquote {\bibinfo {title} {{Beam-Target Helicity Asymmetry for
  $\vec{\gamma} \vec{n} \rightarrow \pi^- p$ in the $N^*$ Resonance Region}},}\
  }\href {\doibase 10.1103/PhysRevLett.118.242002} {\bibfield  {journal}
  {\bibinfo  {journal} {Phys. Rev. Lett.}\ }\textbf {\bibinfo {volume} {118}},\
  \bibinfo {pages} {242002} (\bibinfo {year} {2017})}\BibitemShut {NoStop}%
\bibitem [{\citenamefont {Grames}\ \emph {et~al.}(2004)\citenamefont {Grames}
  \emph {et~al.}}]{Moller2}%
  \BibitemOpen
  \bibfield  {author} {\bibinfo {author} {\bibfnamefont {J.~M.}\ \bibnamefont
  {Grames}} \emph {et~al.},\ }\bibfield  {title} {\enquote {\bibinfo {title}
  {Unique electron polarimeter analyzing power comparison and precision
  spin-based energy measurement},}\ }\href {\doibase
  10.1103/PhysRevSTAB.7.042802} {\bibfield  {journal} {\bibinfo  {journal}
  {Phys. Rev. ST Accel. Beams}\ }\textbf {\bibinfo {volume} {7}},\ \bibinfo
  {pages} {042802} (\bibinfo {year} {2004})}\BibitemShut {NoStop}%
\bibitem [{\citenamefont {Sober}\ \emph {et~al.}(2000)\citenamefont {Sober}
  \emph {et~al.}}]{Sober}%
  \BibitemOpen
  \bibfield  {author} {\bibinfo {author} {\bibfnamefont {D.~I.}\ \bibnamefont
  {Sober}} \emph {et~al.},\ }\bibfield  {title} {\enquote {\bibinfo {title}
  {{The bremsstrahlung tagged photon beam in Hall B at JLab}},}\ }\href
  {\doibase 10.1016/S0168-9002(99)00784-6} {\bibfield  {journal} {\bibinfo
  {journal} {Nucl. Instrum. Meth.}\ }\textbf {\bibinfo {volume} {A440}},\
  \bibinfo {pages} {263--284} (\bibinfo {year} {2000})}\BibitemShut {NoStop}%
\bibitem [{\citenamefont {Olsen}\ and\ \citenamefont
  {Maximon}(1959)}]{Olsen:1959zz}%
  \BibitemOpen
  \bibfield  {author} {\bibinfo {author} {\bibfnamefont {H.}~\bibnamefont
  {Olsen}}\ and\ \bibinfo {author} {\bibfnamefont {L.~C.}\ \bibnamefont
  {Maximon}},\ }\bibfield  {title} {\enquote {\bibinfo {title} {{Photon and
  Electron Polarization in High-Energy Bremsstrahlung and Pair Production with
  Screening}},}\ }\href {\doibase 10.1103/PhysRev.114.887} {\bibfield
  {journal} {\bibinfo  {journal} {Phys. Rev.}\ }\textbf {\bibinfo {volume}
  {114}},\ \bibinfo {pages} {887--904} (\bibinfo {year} {1959})}\BibitemShut
  {NoStop}%
\bibitem [{\citenamefont {Lowry}\ \emph {et~al.}(2016)\citenamefont {Lowry}
  \emph {et~al.}}]{Lowry:2016uwa}%
  \BibitemOpen
  \bibfield  {author} {\bibinfo {author} {\bibfnamefont {M.~M.}\ \bibnamefont
  {Lowry}} \emph {et~al.},\ }\bibfield  {title} {\enquote {\bibinfo {title} {{A
  cryostat to hold frozen-spin polarized HD targets in CLAS: HDice-II}},}\
  }\href {\doibase 10.1016/j.nima.2015.12.063} {\bibfield  {journal} {\bibinfo
  {journal} {Nucl. Instrum. Meth.}\ }\textbf {\bibinfo {volume} {A815}},\
  \bibinfo {pages} {31--41} (\bibinfo {year} {2016})}\BibitemShut {NoStop}%
\bibitem [{\citenamefont {Bass}\ \emph {et~al.}(2014)\citenamefont {Bass} \emph
  {et~al.}}]{Bass:2013noa}%
  \BibitemOpen
  \bibfield  {author} {\bibinfo {author} {\bibfnamefont {C.~D.}\ \bibnamefont
  {Bass}} \emph {et~al.},\ }\bibfield  {title} {\enquote {\bibinfo {title} {{A
  portable cryostat for the cold transfer of polarized solid HD targets:
  HDice-I}},}\ }\href {\doibase 10.1016/j.nima.2013.10.056} {\bibfield
  {journal} {\bibinfo  {journal} {Nucl. Instrum. Meth.}\ }\textbf {\bibinfo
  {volume} {A737}},\ \bibinfo {pages} {107--116} (\bibinfo {year}
  {2014})}\BibitemShut {NoStop}%
\bibitem [{\citenamefont {Drucker}\ and\ \citenamefont
  {Cortes}(1996)}]{DruckerCortes}%
  \BibitemOpen
  \bibfield  {author} {\bibinfo {author} {\bibfnamefont {H.}~\bibnamefont
  {Drucker}}\ and\ \bibinfo {author} {\bibfnamefont {C.}~\bibnamefont
  {Cortes}},\ }\bibfield  {title} {\enquote {\bibinfo {title} {Boosting
  decision trees},}\ }in\ \href@noop {} {\emph {\bibinfo {booktitle} {Advances
  in Neural Information Processing Systems}}},\ \bibinfo {series and number}
  {\bibinfo {number} {8}},\ \bibinfo {editor} {edited by\ \bibinfo {editor}
  {\bibfnamefont {D.}~\bibnamefont {Touretzky}}, \bibinfo {editor}
  {\bibfnamefont {M.}~\bibnamefont {Moser}}, \ and\ \bibinfo {editor}
  {\bibfnamefont {M.}~\bibnamefont {Hasselmo}}}\ (\bibinfo  {publisher} {MIT
  Press},\ \bibinfo {address} {Cambridge, MA},\ \bibinfo {year} {1996})\ pp.\
  \bibinfo {pages} {479--485}\BibitemShut {NoStop}%
\bibitem [{\citenamefont {Roe}\ \emph {et~al.}(2005)\citenamefont {Roe},
  \citenamefont {Yang}, \citenamefont {Zhu}, \citenamefont {Liu}, \citenamefont
  {Stancu},\ and\ \citenamefont {McGregor}}]{ROE2005577}%
  \BibitemOpen
  \bibfield  {author} {\bibinfo {author} {\bibfnamefont {B.~P.}\ \bibnamefont
  {Roe}}, \bibinfo {author} {\bibfnamefont {H-J.}\ \bibnamefont {Yang}},
  \bibinfo {author} {\bibfnamefont {J.}~\bibnamefont {Zhu}}, \bibinfo {author}
  {\bibfnamefont {Y.}~\bibnamefont {Liu}}, \bibinfo {author} {\bibfnamefont
  {I.}~\bibnamefont {Stancu}}, \ and\ \bibinfo {author} {\bibfnamefont
  {G.}~\bibnamefont {McGregor}},\ }\bibfield  {title} {\enquote {\bibinfo
  {title} {Boosted decision trees as an alternative to artificial neural
  networks for particle identification},}\ }\href {\doibase
  https://doi.org/10.1016/j.nima.2004.12.018} {\bibfield  {journal} {\bibinfo
  {journal} {Nucl. Instrum. Meth.}\ }\textbf {\bibinfo {volume} {A543}},\
  \bibinfo {pages} {577 -- 584} (\bibinfo {year} {2005})}\BibitemShut {NoStop}%
\bibitem [{\citenamefont {Sharabian}\ \emph {et~al.}(2006)\citenamefont
  {Sharabian} \emph {et~al.}}]{Sharabian}%
  \BibitemOpen
  \bibfield  {author} {\bibinfo {author} {\bibfnamefont {Y.~G.}\ \bibnamefont
  {Sharabian}} \emph {et~al.},\ }\bibfield  {title} {\enquote {\bibinfo {title}
  {{A new highly segmented start counter for the CLAS detector}},}\ }\href
  {\doibase 10.1016/j.nima.2005.10.031} {\bibfield  {journal} {\bibinfo
  {journal} {Nucl. Instrum. Meth.}\ }\textbf {\bibinfo {volume} {A556}},\
  \bibinfo {pages} {246--258} (\bibinfo {year} {2006})}\BibitemShut {NoStop}%
\bibitem [{\citenamefont {Mestayer}\ \emph {et~al.}(2000)\citenamefont
  {Mestayer} \emph {et~al.}}]{Mestayer}%
  \BibitemOpen
  \bibfield  {author} {\bibinfo {author} {\bibfnamefont {M.~D.}\ \bibnamefont
  {Mestayer}} \emph {et~al.},\ }\bibfield  {title} {\enquote {\bibinfo {title}
  {{The CLAS drift chamber system}},}\ }\href {\doibase
  10.1016/S0168-9002(00)00151-0} {\bibfield  {journal} {\bibinfo  {journal}
  {Nucl. Instrum. Meth.}\ }\textbf {\bibinfo {volume} {A449}},\ \bibinfo
  {pages} {81--111} (\bibinfo {year} {2000})}\BibitemShut {NoStop}%
\bibitem [{\citenamefont {Smith}\ \emph {et~al.}(1999)\citenamefont {Smith}
  \emph {et~al.}}]{Smith}%
  \BibitemOpen
  \bibfield  {author} {\bibinfo {author} {\bibfnamefont {E.~S.}\ \bibnamefont
  {Smith}} \emph {et~al.},\ }\bibfield  {title} {\enquote {\bibinfo {title}
  {{The time-of-flight system for CLAS}},}\ }\href {\doibase
  10.1016/S0168-9002(99)00484-2} {\bibfield  {journal} {\bibinfo  {journal}
  {Nucl. Instrum. Meth.}\ }\textbf {\bibinfo {volume} {A432}},\ \bibinfo
  {pages} {265--298} (\bibinfo {year} {1999})}\BibitemShut {NoStop}%
\bibitem [{\citenamefont {Ho}(2015)}]{Ho-thesis}%
  \BibitemOpen
  \bibfield  {author} {\bibinfo {author} {\bibfnamefont {D.}~\bibnamefont
  {Ho}},\ }\bibfield  {title} {\enquote {\bibinfo {title} {{Measurement of the
  $E$ Polarization Observable for $\gamma d \to \pi^- p (p_s) $, $\gamma d \to
  K^0 \Lambda (p_s) $, and $\gamma d \to \pi^+ \pi^- d(0) $ using CLAS g14 Data
  at Jefferson Lab}},}\ }\href@noop {} {\bibfield  {journal} {\bibinfo
  {journal} {Ph.D. Thesis, Carnegie Mellon University}\ } (\bibinfo {year}
  {2015})},\ \bibinfo {note}
  {\url{http://www.jlab.org/Hall-B/general/clas_thesis.html}}\BibitemShut
  {NoStop}%
\bibitem [{\citenamefont {H{\"o}cker}\ \emph {et~al.}(2007)\citenamefont
  {H{\"o}cker}, \citenamefont {Speckmayer}, \citenamefont {Stelzer},
  \citenamefont {Tegenfeldt},\ and\ \citenamefont {Voss}}]{Hocker:2007zz}%
  \BibitemOpen
  \bibfield  {author} {\bibinfo {author} {\bibfnamefont {A.}~\bibnamefont
  {H{\"o}cker}}, \bibinfo {author} {\bibfnamefont {P.}~\bibnamefont
  {Speckmayer}}, \bibinfo {author} {\bibfnamefont {J.}~\bibnamefont {Stelzer}},
  \bibinfo {author} {\bibfnamefont {F.}~\bibnamefont {Tegenfeldt}}, \ and\
  \bibinfo {author} {\bibfnamefont {H.}~\bibnamefont {Voss}},\ }\bibfield
  {title} {\enquote {\bibinfo {title} {{TMVA, toolkit for multivariate data
  analysis with ROOT}},}\ }in\ \href
  {http://cds.cern.ch/record/1099990/files/p184.pdf} {\emph {\bibinfo
  {booktitle} {{Statistical issues for LHC physics. Proceedings, Workshop,
  PHYSTAT-LHC, Geneva, Switzerland, June 27-29, 2007}}}},\ \bibinfo {editor}
  {edited by\ \bibinfo {editor} {\bibfnamefont {H.B.}\ \bibnamefont {Prosper}},
  \bibinfo {editor} {\bibfnamefont {L.}~\bibnamefont {Lyons}}, \ and\ \bibinfo
  {editor} {\bibfnamefont {A.}~\bibnamefont {DeRoeck}}}\ (\bibinfo {year}
  {2007})\ pp.\ \bibinfo {pages} {184--187}\BibitemShut {NoStop}%
\bibitem [{\citenamefont {Cladis}\ \emph {et~al.}(1952)\citenamefont {Cladis},
  \citenamefont {Hess},\ and\ \citenamefont {Moyer}}]{Cladis_PhysRev.87.425}%
  \BibitemOpen
  \bibfield  {author} {\bibinfo {author} {\bibfnamefont {J.~B.}\ \bibnamefont
  {Cladis}}, \bibinfo {author} {\bibfnamefont {W.~N.}\ \bibnamefont {Hess}}, \
  and\ \bibinfo {author} {\bibfnamefont {B.~J.}\ \bibnamefont {Moyer}},\
  }\bibfield  {title} {\enquote {\bibinfo {title} {Nucleon momentum
  distributions in deuterium and carbon inferred from proton scattering},}\
  }\href {\doibase 10.1103/PhysRev.87.425} {\bibfield  {journal} {\bibinfo
  {journal} {Phys. Rev.}\ }\textbf {\bibinfo {volume} {87}},\ \bibinfo {pages}
  {425--433} (\bibinfo {year} {1952})}\BibitemShut {NoStop}%
\bibitem [{\citenamefont {Lamia}\ \emph {et~al.}(2012)\citenamefont {Lamia},
  \citenamefont {La~Cognata}, \citenamefont {Spitaleri}, \citenamefont
  {Irgaziev},\ and\ \citenamefont {Pizzone}}]{Lamia:2012zz}%
  \BibitemOpen
  \bibfield  {author} {\bibinfo {author} {\bibfnamefont {L.}~\bibnamefont
  {Lamia}}, \bibinfo {author} {\bibfnamefont {M.}~\bibnamefont {La~Cognata}},
  \bibinfo {author} {\bibfnamefont {C.}~\bibnamefont {Spitaleri}}, \bibinfo
  {author} {\bibfnamefont {B.}~\bibnamefont {Irgaziev}}, \ and\ \bibinfo
  {author} {\bibfnamefont {R.~G.}\ \bibnamefont {Pizzone}},\ }\bibfield
  {title} {\enquote {\bibinfo {title} {{Influence of the d-state component of
  the deuteron wave function on the application of the Trojan horse method}},}\
  }\href {\doibase 10.1103/PhysRevC.85.025805} {\bibfield  {journal} {\bibinfo
  {journal} {Phys. Rev.}\ }\textbf {\bibinfo {volume} {C85}},\ \bibinfo {pages}
  {025805} (\bibinfo {year} {2012})}\BibitemShut {NoStop}%
\bibitem [{\citenamefont {Ramachandran}\ \emph {et~al.}(1979)\citenamefont
  {Ramachandran}, \citenamefont {Keshavamurthy},\ and\ \citenamefont
  {Murthy}}]{Ramachandran:1979ck}%
  \BibitemOpen
  \bibfield  {author} {\bibinfo {author} {\bibfnamefont {G.}~\bibnamefont
  {Ramachandran}}, \bibinfo {author} {\bibfnamefont {R.~S.}\ \bibnamefont
  {Keshavamurthy}}, \ and\ \bibinfo {author} {\bibfnamefont {M.~V.~N.}\
  \bibnamefont {Murthy}},\ }\bibfield  {title} {\enquote {\bibinfo {title}
  {{Target Asymmetry And Effective Neutron Polarization With Polarized Deuteron
  Targets}},}\ }\href {\doibase 10.1016/0370-2693(79)90977-8} {\bibfield
  {journal} {\bibinfo  {journal} {Phys. Lett.}\ }\textbf {\bibinfo {volume}
  {87B}},\ \bibinfo {pages} {252--256} (\bibinfo {year} {1979})}\BibitemShut
  {NoStop}%
\bibitem [{\citenamefont {Casey}(2011)}]{LiamCasey}%
  \BibitemOpen
  \bibfield  {author} {\bibinfo {author} {\bibfnamefont {L.}~\bibnamefont
  {Casey}},\ }\bibfield  {title} {\enquote {\bibinfo {title} {{The Search for
  Missing Resonances in $\gamma p \to K^+ \Lambda$ Using Circularly Polarized
  Photons on a Longitudinally Polarized Frozen Spin Target}},}\ }\href@noop {}
  {\bibfield  {journal} {\bibinfo  {journal} {Ph.D. Thesis, Catholic University
  of America}\ } (\bibinfo {year} {2011})},\ \bibinfo {note}
  {\url{http://www.jlab.org/Hall-B/general/clas_thesis.html}}\BibitemShut
  {NoStop}%
\end{thebibliography}%
\end{document}